\documentclass[preprint,graphicx]{aastex}

\usepackage{natbib}
\slugcomment{Accepted by \aj}

\lefthead{Rider et~al.}
\righthead{$u'g'r'i'z'$ Photometry of NGC~2548 }

\begin{document}

\title{A Survey of Open Clusters in the $u'g'r'i'z'$
Filter System: I. Results for NGC~2548 (M~48)}

\author{
Cristin J. Rider\altaffilmark{\ref{BU},\ref{Fermilab},3,4},
Douglas L. Tucker\altaffilmark{\ref{Fermilab},5},
J. Allyn Smith\altaffilmark{5,6,7,8},
Chris Stoughton\altaffilmark{\ref{Fermilab}},
Sahar S. Allam\altaffilmark{\ref{Fermilab},5,9},
Eric H. Neilsen Jr.\altaffilmark{\ref{Fermilab}}
}

\altaffiltext{1}{Boston University,
            Astronomy Department, Boston University,
            725 Commonwealth Avenue, Boston, MA  02215
  \label{BU}}
\altaffiltext{2}{Fermi National Accelerator Laboratory,
                 P.O. Box 500, Batavia, IL 60510
  \label{Fermilab}}
\altaffiltext{3}{Fermi National Accelerator Laboratory Internship
            for Undergraduate Physics Majors program.}
\altaffiltext{4}{Current address Johns Hopkins University, 
                 Department of Physics \& Astronomy, Homewood Campus, 
                 3701 San Martin Drive, Baltimore, MD  21218 }

\altaffiltext{5}{Visiting Astronomer, US Naval Observatory, Flagstaff, AZ, 86002}
\altaffiltext{6}{Los Alamos National Laboratory, NIS-4, D448,
            Los Alamos, NM 87545}
\altaffiltext{7}{University of Wyoming,
            Department of Physics \& Astronomy,
            P.O. Box 3905,
            Laramie, WY 82071}
  \label{Wyoming}
\altaffiltext{8}{Visiting Astronomer, Cerro Tololo Inter-American
            Observatory.  CTIO is operated by AURA, Inc. under contract to
            the National Science Foundation.}
\altaffiltext{9}{New Mexico State University, Astronomy Department, 
                 Box 30001, Dept 4500, 1320 Frenger St., Las Cruces NM 88003}

\begin{abstract}
We  present  initial  results of  a  photometric survey   of open star
clusters,   primarily  in the    southern hemisphere,   taken   in the
$u'g'r'i'z'$ filter system.   While our entire  observed sample covers
more than 100 clusters, here we present data for NGC~2548 (M~48) which
is a cluster characterized in the $UBV$ and $DDO$ photometric systems.
We compare our results  to the published  values from  other observers
and   to the Padova    theoretical isochrones and  metallicity curves.
These observations demonstrate  that the $u'g'r'i'z'$ filters can play
an  important  role   in determining  the  metallicity   of stars  and
clusters.  We begin this  series of  papers  with a study of  NGC~2548
because we have obtained data  of this cluster not  only with our main
program telescope, the CTIO Curtis-Schmidt, but also with the US Naval
Observatory (USNO)  1.0~m telescope (the  telescope used to define the
$u'g'r'i'z'$ system), and  the Sloan Digital  Sky  Survey (SDSS) 0.5~m
Photometric Telescope  (the  photometric monitoring telescope  used to
calibrate  the SDSS  2.5~m telescope  imaging data). We  have used the
data from this study to validate our ability to transform measurements
obtained on other telescopes  to the standard USNO~1.0~m  $u'g'r'i'z'$
system.  This validation is particularly important for very red stars,
for  which the original  $u'g'r'i'z'$  standard star network is poorly
constrained.
\end{abstract}


\section{Introduction}

One of the most powerful tools available in astrophysics for exploring
and   testing theories   of   star formation,   stellar  and  galactic
evolution, and the chemical enrichment  history of  the Galaxy is  the
study of open star clusters.  Because of this, open clusters have been
the subject  of intense  studies  for the past several  decades (e.g.,
\citealt{ATTM79,GS84,ATTS94,BFKA99,WOCS1}; see also \citealt{Friel95}
and references therein).

We have  embarked  on a survey  of (mostly)  southern  hemisphere star
clusters  using the    $u'g'r'i'z'$   filter  system.  The    original
motivation  of  the project was  to use  these clusters, which  span a
range  of ages  and metallicities,  to  ``back  calibrate'' the  Sloan
Digital Sky Survey (SDSS) [see \citet{GCRS98} for a description of the
SDSS survey camera and \citet{york00} for  a description of the SDSS].
In addition, these data  may now be used  to verify the recent age and
metallicity models of \citet{Girardi03} and the prior work of
\citet{Lenz} and to verify and expand upon the $u'g'r'i'z'$ to $UBVRI$
transformations presented in the standard star paper \citep{Smith02}.

The  initial  effort  in    this  survey obtained   observations   for
approximately 105 open  clusters  and  a few ($<$10)   ``low-density''
globular clusters.  This approach was driven, in  part, as a result of
the  availability of the Curtis-Schmidt  telescope at the Cerro Tololo
Inter-American Observatory (CTIO).  This  telescope was available as a
result  of one of us  (JAS) being at  the University  of Michigan as a
research fellow.  The need  to perform the  survey observations was  a
result  of the impending closure  of this  wonderful survey telescope.
The wide   field  of this instrument allowed    us  to survey  several
clusters over the course of four observing runs, from 1997--2000.

In this, the first paper  of our  series,  we present our results  for
NGC~2548,  also  known  as M~48, which   is  an intermediate  age open
cluster. We chose this  particular cluster because we had observations
of it  from three telescopes:   the CTIO Curtis-Schmidt, which  is the
main telescope  for our open clusters   program; the 0.5~m Photometric
Telescope, which is  used to calibrate the imaging  data from the SDSS
2.5~m telescope; and, most    importantly, the US   Naval  Observatory
(USNO)  1.0~m  telescope  at  Flagstaff  Station,   which was used  to
establish the original $u'g'r'i'z'$ standard star network
\citep{Smith02}.   These   data, especially  from    the  USNO  1.0~m
telescope, have allowed us to tie  the CTIO Curtis-Schmidt system more
strongly to  the basic definition  of the  $u'g'r'i'z'$  system.  As a
result the  findings from the remainder of  our  open clusters program
will be placed on a firmer foundation.

Although NGC~2548 is bright enough  to have been cataloged by Messier,
it   was  once thought  to be   non-existent due  to   a sign error in
Messier's reported coordinates \citep{Wu02}.   While this cluster  has
been   the  subject   of      several    proper  motion        studies
\citep{Ebbighausen39,Dias01,Wu02}, chemical composition studies 
\citep{WalC64,Claria85,Gilroy89}, and dynamical studies
\citep{Geyer85,BLG01}, there have been surprisingly few photometric
studies \citep{Pesch61,Claria85}.  The only  photometric study of this
cluster consisting of more than five stars  is the $UBV$ photoelectric
observations obtained by   \citet{Pesch61}, who observed 37 stars  and
divided  them   into spectral  groups.    Further,  he  estimated  the
reddening to be $E(B-V)=.04\pm.05$ and the distance to be 630~pc.

In  the following sections we present   details of the instrumentation
and  observations,  (\S2),   data reduction  and analysis  techniques,
(\S3), results (\S4), and planned future work (\S5).

\section{Instrumentation and Observations}

\subsection{$u'g'r'i'z'$ Filter System}

The five filters of the $u'g'r'i'z'$ system have effective wavelengths
of 3540\AA, 4750\AA,  6222\AA, 7632\AA, and  9049\AA, respectively, at
1.2 airmasses.\footnote{Note that the  $g'$ filter has been determined
to have  an effective wavelength 20  \AA\, bluer  than that originally
quoted by \citet{fuk96}.}   They cover the  entire wavelength range of
the  combined   atmosphere+CCD  response  and  their  construction  is
described  by   \citet{fuk96}.  The $u'g'r'i'z'$  filters  have  sharp
cutoffs by  design.  The   passbands   were selected  to  exclude  the
strongest night-sky lines; for example \ion{O}{1} ($\lambda$5577) and
\ion{Hg}{1}  ($\lambda$5461).  The bulk of   the $u'$ band response is
blueward of the  Balmer discontinuity  which,  when combined with  the
$g'$ filter, yields  high sensitivity to the  magnitude  of the Balmer
jump but at a  cost of lower  throughput for the narrower  $u'$ filter
(compared with Johnson $U$).  We note the red side of the $z'$ filter
is open ended, and therefore subject to potentially large color terms
in the transformation equations, dependent upon the choice of detector.

The  $u'g'r'i'z'$ magnitude system is    a broadband $AB$ system.   In
other words, rather than selecting some fiducial spectral type to have
null colors (as in the $UBVR_{\rm c}I_{\rm c}$ system), $AB$ broadband
magnitudes are defined by the following equation:
\begin{equation}
m = -2.5\log\frac{ \int d(\log \nu) f_{\nu} S_{\nu} }{ \int d(\log \nu) S_{\nu}} - 48.60 , \label{eq:broadbandAB}
\end{equation}
where  $f_{\nu}$   is  the energy flux   per  unit  frequency  on  the
atmosphere and $S_{\nu}$  is the  system response.   Hence in an  $AB$
system, an object with a  flat spectral energy distribution would have
null colors \citep{OG83,fuk96}.

To   zeropoint  the $u'g'r'i'z'$   system,  \citet{Smith02}   used the
synthetic   $AB$  $u'g'r'i'z'$    magnitudes    of the    F   subdwarf
BD+17$\arcdeg$4708  as calculated by \citet{fuk96}.  Initial estimates
indicate that  the $u'g'r'i'z'$  network  is offset  from a true  $AB$
broadband system by no more than about 5\% in $u'$  and $z'$ and by no
more than about 3\% in $g'r'i'$ (see \S~4 below), due to uncertainties
in the absolute calibration  of the synthetic $u'g'r'i'z'$  magnitudes
of BD+17$\arcdeg$4708.

The   $u'g'r'i'z'$ standard star    network is composed  of  158 stars
distributed  primarily along the  celestial  equator and the  northern
celestial    hemisphere \citep{Smith02}. As     a cautionary note, the
primary goals  for the SDSS  are  large scale structure  studies using
galaxies and QSOs, so  the  first version  of the standard  network is
limited, for the most part, to stars bluer than about  M0 to avoid the
strengthening metal bands and flare stars.

\subsection{ Telescopes and Observations }

As noted above, we  are using data from  three separate  telescopes in
this study of NGC~2548: the  USNO 1.0~m telescope at Flagstaff Station
in Arizona, the SDSS 0.5~m Photometric  Telescope (PT) at Apache Point
Observatory in New Mexico,  and the University of Michigan's 0.6/0.9~m
Curtis-Schmidt telescope at CTIO (CTIO-CS) in Chile.  We will describe
each instrument   in  turn.  A  summary  of the  system  parameters is
presented  in  Table~\ref{systems}.     A  summary of  the   observing
circumstances for the NGC~2548 field is given in Table~\ref{obs}.

%Table 1 - system parameters summary
\placetable{systems}
\placetable{obs}

\subsubsection{USNO 1.0~m }

The USNO 1.0~m  telescope  is  of Ritchey--Chr\'{e}tien  design.   The
observations for our NGC~2548 program on  this telescope were obtained
on  2002 November 5  and 6 (UT).  All  of the observations were direct
exposures with a thinned, UV-AR coated, Tektronix TK1024 CCD operating
at a gain of 7.43$\pm$0.41 electrons  per ADU with  a readnoise of 6.0
electrons.  This CCD is   similar to the CCDs  used  in the main  SDSS
survey camera and the  CCD used by  the 0.5-m Photometric Telescope at
APO.  This is  the  detector which  defines  the $u'g'r'i'z'$ standard
star  network.   The  camera   scale  of 0.68  arcsec/pixel   for this
$1024\times1024$ detector produces a field of view of 11.54~arcmin.

During a typical night at the telescope, we generally observed four to
five standard   fields at  the start  of  the  night to  determine the
extinction.  Following   this, we would usually   observe two to three
target fields, and then would alternate  between two to three standard
and target fields through the remainder of the night, finishing with a
longer run of standards (usually four-five).  In general, two or three
standard  fields were  observed several  times each  night to  monitor
extinction manually  at the telescope and to  look for changes  in the
photometricity of  the  sky.   These  values  were  compared  with the
``all--sky'' extinction values determined  later during  the reduction
process.  Additional fields  were observed throughout  the night, near
the meridian and at  high airmass, in order   to provide a  good color
spread to solve for instrumental color terms.

\subsubsection{Photometric Telescope}

The SDSS  0.5~m PT  is  a classic  Cassegrain  with an  additional two
corrective   lenses.  The addition   of   these two corrective  lenses
increased the field of view to about 43~arcmin and changed the f-ratio
from f/8.0 to f/8.8.   The CCD is a  SITe $2048\times2048$ device with
24 micron (1.15~arcsec) pixels and a UV-AR coating.

The PT observations  for our  NGC~2548  program were obtained on  2001
December  2 (UT).  On this  night, observing began by targetting three
standard   star  fields covering  a   range  of airmasses and  colors.
Similar  sequences  of observations  occurred   at roughly  90  minute
intervals through the night, which ended with sequence of observations
of four standard star fields, providing data approprate for monitoring
the stability  of the extinction  through the  night and generating an
all-sky photometric solution for the night.

\subsubsection{CTIO Curtis-Schmidt}

The CTIO-CS is  a f/3.5 modified Newtonian focus  system with a  0.9~m
primary mirror and 0.6~m corrector lens.  The CTIO-CS observations for
our  NGC~2548 program were  obtained 2000  March 6  (UT).  During this
observing  session,  we   used the Tek2k\#5    (Tektronix TK2048  CCD)
detector  with the   CTIO   set of  SDSS  $u'g'r'i'z'$  filters.   The
$2048\times2048$ detector has a scale of 2.3~arcsec/pixel, yielding an
effective field of view of 1.3~deg.  Readout of the chip was driven by
two amplifiers and the CCD system was operated under the Arcon control
software (version 3.17) at   a gain setting   \#2.  This gain  setting
yielded  effective   gain    and  read   noise values   as    given in
Table~\ref{systems}.   For  the CTIO-CS observations,  we  followed an
observing  plan    similar  to  that described     for  the USNO~1.0~m
observations ({\S2.2.1}).

\section{Data Reduction}

We performed reductions using version  {\tt v8.0} of the SDSS software
pipeline {\tt  mtpipe} \citep[see][]{Tucker03}.   This version of {\tt
mtpipe} is   being used  for    the  current  PT  reductions used   in
calibrating the SDSS  imaging data \citep[e.g.,][]{abazajian03} and in
the  CTIO-0.9~m  reductions     used for  establishing   a    southern
$u'g'r'i'z'$  standard  star  network    \citep{Smith03}\footnote{{\tt
http://home.fnal.gov/$\sim$dtucker/Southern\_ugriz/index.html}}.    An
earlier version  of {\tt mtpipe}  ({\tt v6.6}) was used for processing
the data   for the setup  of the   original $u'g'r'i'z'$ standard star
network \citep{Smith02}.  (The main changes  between {\tt mtpipe v6.6}
and {\tt v8.0}  lie in the latter  version's increased support for the
PT  and the   CTIO  telescopes and its   use   of an  improved  set of
photometric equations.)   

The {\tt mtpipe} pipeline  consists of four  main packages:
\begin{itemize}

\item {\tt preMtFrames}, which creates the directory structure for the
	reduction of a night's  data, including parameter files needed
	as  input  for     the    other three packages,     and   runs
	quality-assurance tests on the raw data.

\item {\tt mtFrames}, which processes the images and performs object
	detection and aperture photometry on target field images.  The
	processing steps  include  bias (zero) subtraction, flat-field
	and fringe-frame correction.

\item {\tt excal}, which takes the aperture photometry lists for the
	standard    star   target        fields \citep[i.e.,     stars
	from][]{Smith02},  identifies   the individual  standard stars
	within  those fields, and   fits  the observed raw  counts and
	known   $u'g'r'i'z'$ magnitudes  to   a  set  of   photometric
	equations to obtain   extinction and zero  point coefficients.
	The  output   from  this package   allows  us to   monitor the
	stability of the  night.  The default  analysis block is three
	hours,  but  can be  changed as  required  based upon the data
	present  and upon   trends  in the   reductions. Typically,  a
	minimum of ten standard stars are required to obtain a good 
        solution for the night.  

\item {\tt kali}, which applies the fitted photometric equations to
	the aperture photometry lists of program target fields for the
	appropriate analysis block (e.g., the NGC~2548 field).

\end{itemize}

The  photometric equations employed in  {\tt  excal}  for the  current
paper are the following:

\begin{eqnarray}
u'_{\rm inst} & = & u'_{\rm o} + a_{u} + k_{u} X \nonumber \\
               &   & + b_{u} [(u'-g')_{\rm o} - (u'-g')_{\rm o,zp}] \nonumber \\
               &   & + c_{u} [(u'-g')_{\rm o} - (u'-g')_{\rm o,zp}] [X-X_{\rm zp}]\;, \\
g'_{\rm inst} & = & g'_{\rm o} + a_{g} + k_{g} X \nonumber \\
               &   & + b_{g} [(g'-r')_{\rm o} - (g'-r')_{\rm o,zp}] \nonumber \\
               &   & + c_{g} [(g'-r')_{\rm o} - (g'-r')_{\rm o,zp}] [X-X_{\rm zp}]\;, \\
r'_{\rm inst} & = & r'_{\rm o} + a_{r} + k_{r} X \nonumber \\
               &   & + b_{r} [(r'-i')_{\rm o} - (r'-i')_{\rm o,zp}] \nonumber \\
               &   & + c_{r} [(r'-i')_{\rm o} - (r'-i')_{\rm o,zp}] [X-X_{\rm zp}]\;, \\
i'_{\rm inst} & = & i'_{\rm o} + a_{i} + k_{i} X \nonumber \\
               &   & + b_{i} [(i'-z')_{\rm o} - (i'-z')_{\rm o,zp}] \nonumber \\
               &   & + c_{i} [(i'-z')_{\rm o} - (i'-z')_{\rm o,zp}] [X-X_{\rm zp}]\;, \\
z'_{\rm inst} & = & z'_{\rm o} + a_{z} + k_{z} X \nonumber \\
               &   & + b_{z} [(i'-z')_{\rm o} - (i'-z')_{\rm o,zp}] \nonumber \\
               &   & + c_{z} [(i'-z')_{\rm o} - (i'-z')_{\rm o,zp}] [X-X_{\rm zp}] .
\end{eqnarray}

Taking the $g'$ equation  as an example,  we note that $g'_{\rm inst}$
is   the measured  instrumental    magnitude,  $g'_{\rm  o}$  is   the
extra-atmospheric   magnitude,   $(g'-r')_{\rm   o}$   is          the
extra-atmospheric color, $a_{g}$ is the nightly zero point, $k_{g}$ is
the first     order  extinction coefficient,  $b_{g}$   is  the system
transform coefficient, $c_{g}$ is  the second order (color) extinction
coefficient, and $X$ is the airmass of the observation.  The zeropoint
constants,   $X_{\rm  zp}$   and  $(g'-r')_{\rm  o,zp}$  were defined,
respectively, to  be  the average  standard  star observation  airmass
$<X>$  = 1.3  and the  ``cosmic  color,''  as  listed  in Table~3   of
\citet{Smith02}.

In  general we  follow  the reduction procedures  outlined in  \S~3 of
\citet{Smith03}, with the following exceptions:

\noindent{{\bf USNO~1.0~m:}} Since this combination of telescope$+$detector$+$filters
{\em defines\/} the $u'g'r'i'z'$ photometric  system, we do not  solve
for the  coefficient of the   the  intrumenal color terms  (the  ``b''
terms), but set them  all    identically to  zero.  Further, due    to
variations in seeing during the   nights we observed NGC~2548 on  this
telescope,  we found  that  a  larger  aperture size yielded   stabler
aperture photometry and  improved   residuals  in  the  fits  to   the
photometric equations;  so,  instead   of  the  14.86~arcsec  aperture
diameter used in   \citet{Smith03}, we used  the 24.0~arcsec  aperture
diameter originally employed in the setup of the $u'g'r'i'z'$ standard
star network \citep{Smith02}.

\noindent{{\bf Photometric Telescope:}}  The data from the PT for this  
night was originally reduced as part  of the standard SDSS photometric
calibration program.  Currently,    as  part of   this  program,  {\tt
mtFrames}  performs  aperture photometry  using two  default  aperture
sizes: a  large aperture   of  24.0~arcsec and  a small   aperture  of
12.0~arcsec  diameter.   The  large  aperture  size  is used    on the
$u'g'r'i'z'$ standard star target  fields.  The small aperture size is
used on the program target fields.  A correction  to convert the small
aperture  counts to large   aperture counts is  calculated for program
target fields on an image-by-image basis by measuring the ratio of the
fluxes in both apertures for the  12 brightest stars  in the image and
taking  a clipped  mean;  sanity   checks  prevent  the   inclusion of
saturated stars,  cosmic  rays,   or close   optical  doubles  in this
calculation.  The resulting aperture  correction is applied to all the
stars in the target field.

We  took  the original  {\tt  mtFrames}  aperture photometry  for this
night's data and re-ran {\tt excal} and {\tt  kali} on it.  We did not
solve for the  $b$ term coefficients  but set them to the site-average
values   for   the   current    PT   $u'g'r'i'z'$  filter     set (see
Table~\ref{nightChar}   below).   Further,   rather  than  solve   the
photometric equations in 3-hour blocks,  we followed the standard SDSS
convention   used in {\tt mtpipe}    processing and used a  full-night
block.

Note that, in  this  paper, we follow   a slightly different  tack  in
calibrating the PT data than is  done for normal  SDSS operations.  In
particular, in normal SDSS  operations, different values are used  for
the zeropoint colors and in the photometric  equations the $i'$ filter
is indexed to $r'-i'$, not to $i'-z'$ (as  in eq.~[5] above).  We have
compared  the  results from both PT   calibration  strategies for this
night's data, and  find  typical offsets in the  calibrated magnitudes
for any given star to be $\ll$ 0.01~mag.   (For a detailed description
of  the    PT calibration  strategy in   normal   SDSS operations, see
\citealt{Tucker03}.)

\noindent{{\bf CTIO Curtis-Schmidt:}}  To tighten the scatter in
the residuals,  we  chose to  fit  the  photometric equations in  {\tt
excal} using  a single,  full-night  block instead of  multiple 3-hour
blocks.  Other than  that, for  this  telescope,  we follow the   data
reduction methodology of \citet{Smith03} very closely.

The night characterization  data  for each  of the  photometric nights
included in this study of NGC~2548 are given in Table~\ref{nightChar}.
These data include   the Modified Julian Date\footnote   {The Modified
Julian Date  is  defined by the relation  MJD$\equiv$JD$-$2,400,000.5,
where JD is the Julian Date.}  of the  observation (col.  [1]), filter
(col.  [2]), zero   points  (col.  [3]), system   transformation terms
(col.    [4]), and   first-order extinction  terms   (col.  [5]--[7]).
Finally, columns (8) and (9) give the rms errors  for, and numbers of,
the  standard  stars  observed  that night  which   were  used  in the
photometric solutions.  In  a footnote, we  also list the second-order
extinction terms derived in \citet{Smith02}.

\placetable{nightChar}

Once we had  run all the nights  containing NGC~2548 data through {\tt
mtpipe},  we   had  ten  lists   of  calibrated  NGC~2548 $u'g'r'i'z'$
photometry ---  one  for  each   targetting   of this  cluster    (see
Table~\ref{obs}).   In  order to  compare cross-calibrations, we first
combined the contents of these lists by telescope.  In other words, we
combined the contents of the  seven USNO~1.0~m lists together into one
large USNO~1.0m list, and we likewise combined the contents of the two
CTIO-CS lists into a  single  large CTIO-CS  list..  (Since there  was
only   one PT list  to  begin with,  its  contents did not  need to be
combined.)  We did  the combining   by  assigning  each star  in   the
combined list the ``best'' magnitude of its corresponding entries from
all $N$ lists for that telescope.  In this context, ``best'' refers to
the magnitude that has  the smallest photon  noise.  This is done on a
filter-by-filter basis, so a star's  best $u'$ magnitude and best $r'$
magnitude may come from  different lists.  (Star entries were  matched
by position between  lists using a $\pm$2~arcsec box  in RA  and DEC.)
Magnitude entries which have been  flagged by {\tt mtFrames} as  being
saturated or which have  poorly determined values (magnitudes $<0$  or
$>100$) were excluded from the combine procedure.

To reduce field star contamination only stars  within a cluster radius
of  27~arcmin \citep{Lynga87}  were  included  in the  final telescope
lists.  Where possible,  a cluster  membership probability based  upon
\citet{Dias01}'s proper motion study was assigned to each star.

The full data tables  for each telescope are available in the electronic edition of this paper.  The
available  data for  each  of the telescopes  include  our internal ID
number for each  star (col. [1]), the star  ID number from Jean-Claude
Mermilliod's {\tt  webda} online open  cluster  database\footnote{{\tt
http://obswww.unige.ch/webda/webda.html}} (col.   [2]),  RA (col [3]),
DEC (col. [4]), $ u'g'r'i'z'$ magnitudes (cols. [5], [6], [7], [8], \&
[9], respectively),  $  u'g'r'i'z'$ magnitude  (photon  noise)  errors
(cols. [10],  [11],      [12], [13], \&   [14],    respectively),  and
$u'g'r'i'z'$ saturation flags (cols.  [15], [16], [17], [18], \& [19],
respectively).  Lastly  column  (20) gives   the stars  proper  motion
membership  probability from  \citet{Dias01}.   Tables~\ref{dataUSNO},
~\ref{dataPT}, and ~\ref{dataCTIO} show   the first 50 entries  of the
available  data  for the  USNO   1.0 m,   PT, and  CTIO Curtis-Schmidt
telescopes, respectively.

We  also created  a master list  in  which the  lists  from  all three
telescopes were combined.   We   base our internal   numbering  scheme
(col. [1] of  Tables~\ref{dataUSNO}, \ref{dataPT}, and \ref{dataCTIO})
on this final, master list.

\placetable{dataUSNO} 
\placetable{dataPT}
\placetable{dataCTIO}

\section{Results}

\subsection{ System Verification}

\begin{figure}
\includegraphics[angle=0,scale=0.8]{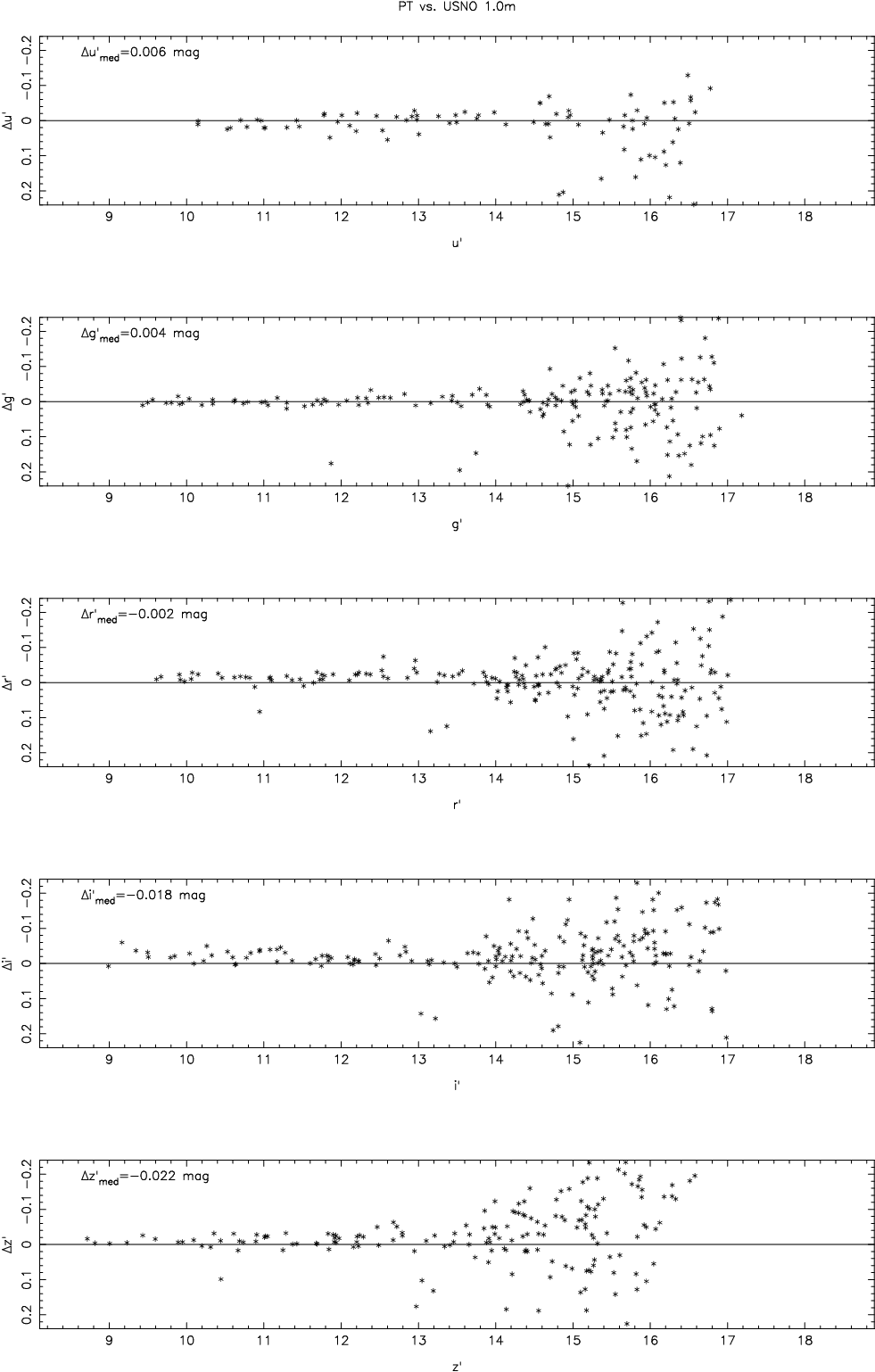}
\caption{ Differences between the USNO 1.0~m and the PT  magnitudes 
as a function of apparent magnitude for each of the five filters.  The
value  in the top  left corner of each panel  is  the median magnitude
offset between the USNO~1.0~m measurements  and the PT measurements in
that filter.
\label{usnoPTvsMag}}
\end{figure}

\begin{figure}
\includegraphics[angle=0,scale=0.8]{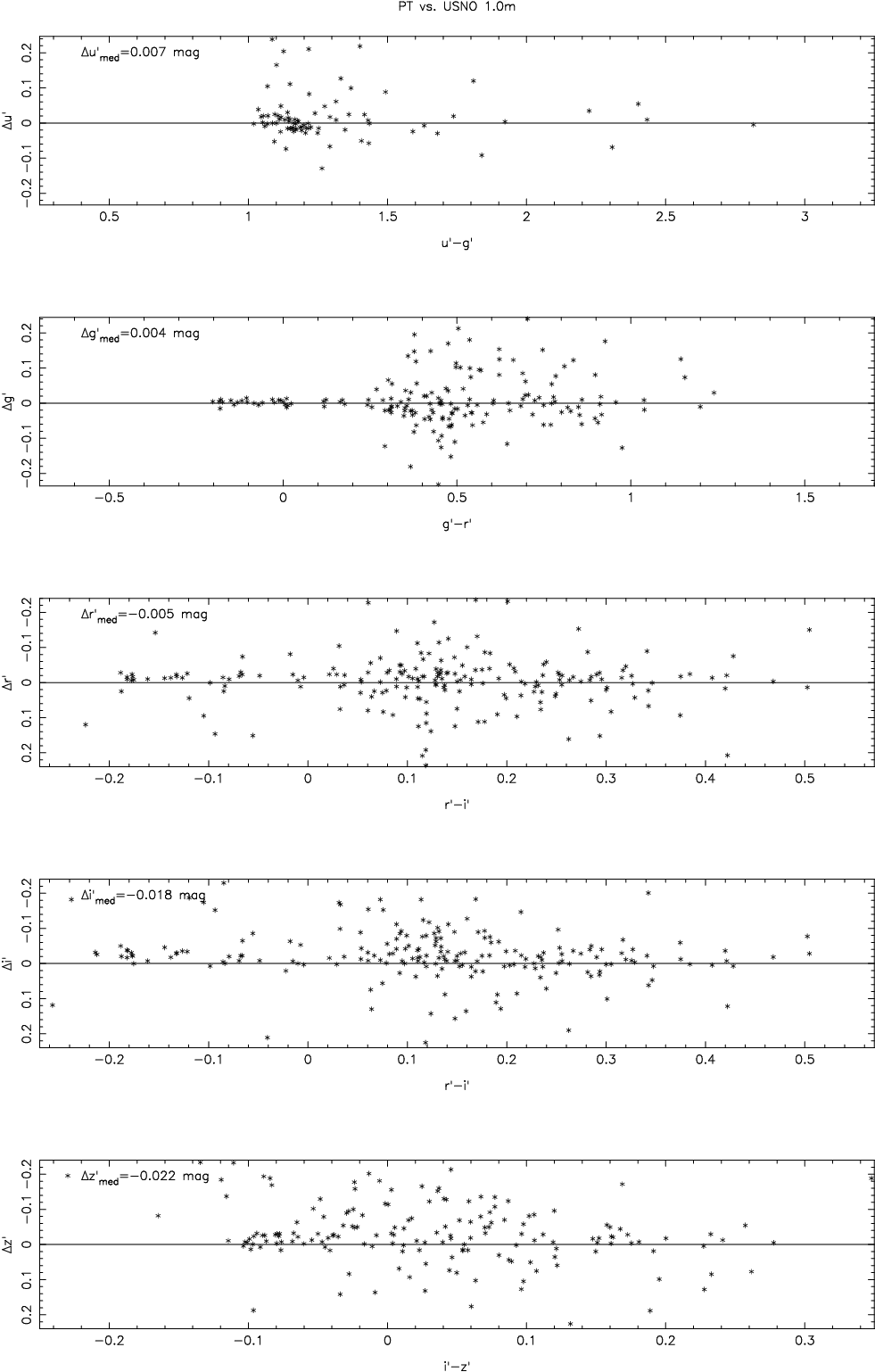}
\caption{ Differences between the USNO 1.0~m and the PT  magnitudes 
as a function of color for each of the five filters.  The value in the
top left corner of  each panel is  the median magnitude offset between
the USNO~1.0~m measurements and the PT measurements in that filter.
\label{usnoPTvsColor}}
\end{figure}

\begin{figure}
\includegraphics[angle=0,scale=0.8]{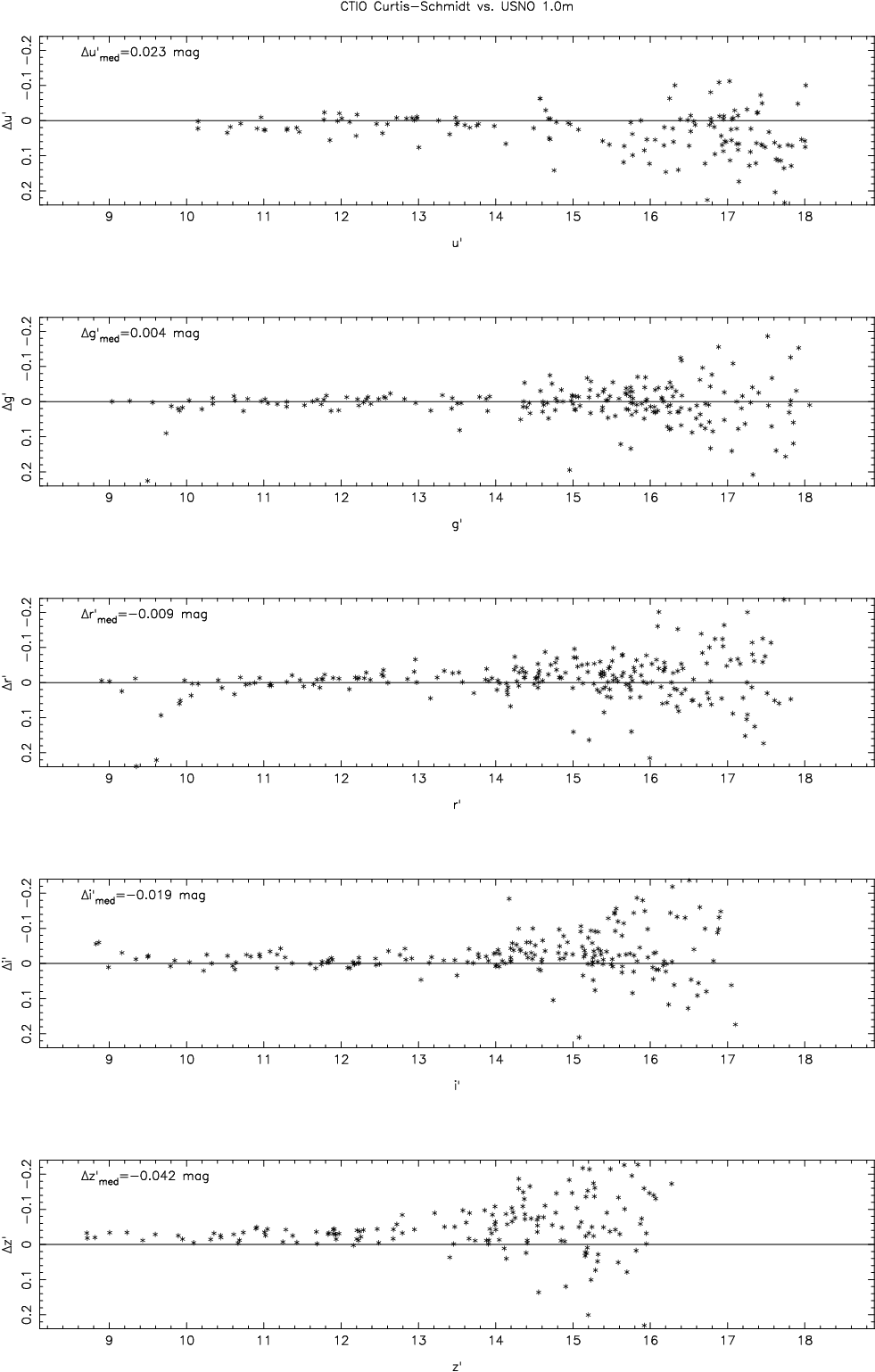}
\caption{ Differences between the USNO 1.0~m and the CTIO-CS  magnitudes 
as a function of apparent magnitude for each of the five filters.  The
value  in the top  left corner of each panel  is  the median magnitude
offset between the USNO~1.0~m measurements  and the CTIO-CS measurements in
that filter.
\label{usnoCTIOvsMag}}
\end{figure}

\begin{figure}
\includegraphics[angle=0,scale=0.8]{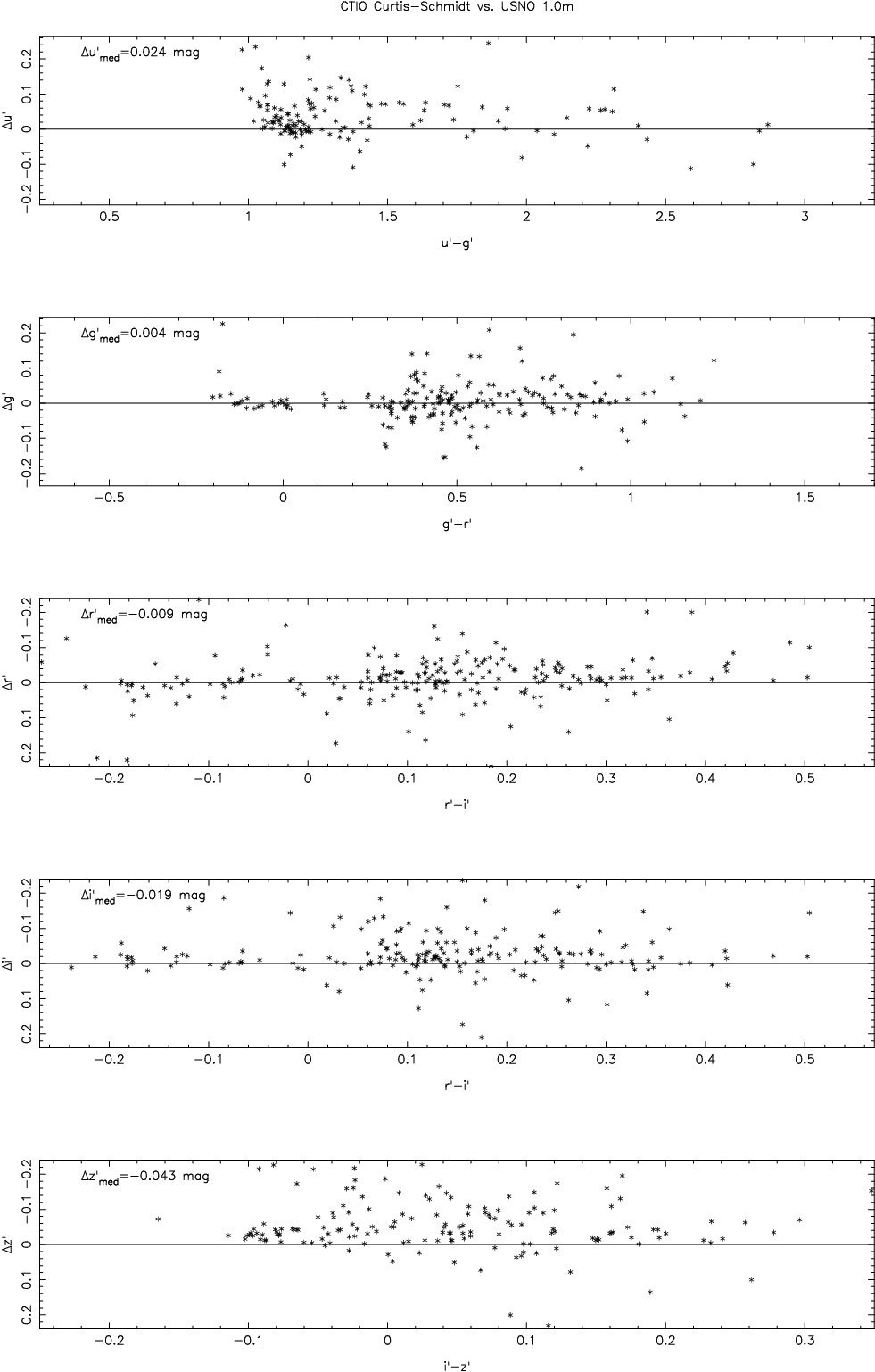}
\caption{ Differences between the USNO 1.0~m and the CTIO-CS  magnitudes 
as a function of apparent color for each of the five filters.  The
value  in the top  left corner of each panel  is  the median magnitude
offset between the USNO~1.0~m measurements  and the CTIO-CS measurements in
that filter.
\label{usnoCTIOvsColor}}
\end{figure}

The USNO 1.0~m telescope defines  the $u'g'r'i'z'$ filter system.   We
verify   our ability  to apply   this  system  to  other telescopes by
comparing our observations from  those  telescopes to those  from  the
USNO 1.0~m telescope.  This was done by  determining the difference in
the   magnitude, for  each filter, for    all stars  observed by  both
telescopes.   Figures~\ref{usnoPTvsMag} and \ref{usnoPTvsColor}  show,
respectively, the differences in  magnitude as a function of magnitude
and as a  function of color between  the USNO 1.0~m  telescope and the
PT.   Likewise,  Figures~\ref{usnoCTIOvsMag} and \ref{usnoCTIOvsColor}
show,  respectively,  the differences  in  magnitude as a  function of
magnitude and as a function of color  between the USNO 1.0~m telescope
and the CTIO Curtis-Schmidt   telescope.   As measured by the   median
magnitude     offsets    in each   filter,  there     is  a $\la$2.2\%
($\leq$0.022~mag) systematic difference between the USNO 1.0~m and the
PT  magnitudes and ---   except for  the  $z'$  band ---  a $\la$2.4\%
($\leq$0.024~mag) systematic difference between the USNO 1.0~m and the
CTIO         Curtis-Schmidt      magnitudes.              Furthermore,
Figures~\ref{usnoPTvsMag},   \ref{usnoPTvsColor},  \ref{usnoCTIOvsMag}
and \ref{usnoCTIOvsColor} indicate that   there are no strong  overall
trends  in the systematic  differences  as a function  of magnitude or
color for  these telescopes.   The somewhat larger  USNO~1.0~m/CTIO-CS
$z'$ band offset (0.042~mag or about 4.2\%) could  be ascribed to just
random variations  in the atmosphere (note  the relatively high rms in
the fit to the  $z'$ photometric equation   for the this night on  the
CTIO-CS in Table~\ref{nightChar}).  It  may also indicate a difficulty
in   performing $z'$ band  aperture photometry   with this instrument.
This issue will  likely be resolved as we  analyze more of the CTIO-CS
data in detail.

\placefigure{usnoPTvsMag}

\placefigure{usnoPTvsColor}

\placefigure{usnoCTIOvsMag}

\placefigure{usnoCTIOvsColor}

\begin{figure}
\plotone{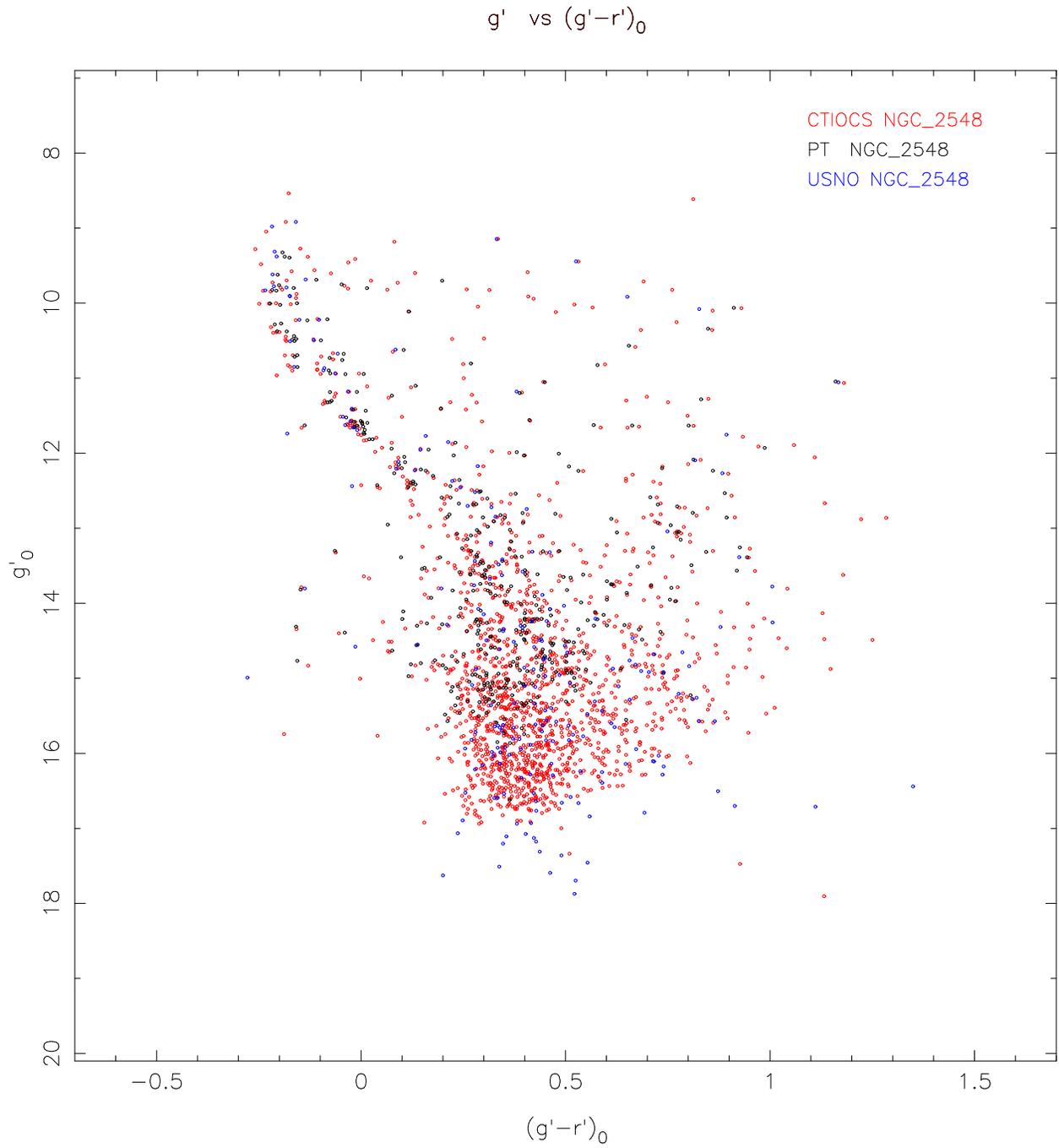}
\caption{A comparison of the de-reddened $g'$ vs. $g'-r'$ color-magnitude diagrams for 
NGC~2548 obtained from the CTIO-CS (red dots), the PT (black dots), and the USNO~1.0m
(blue dots).  
\label{m2548}}
\end{figure}

\begin{figure}
\plotone{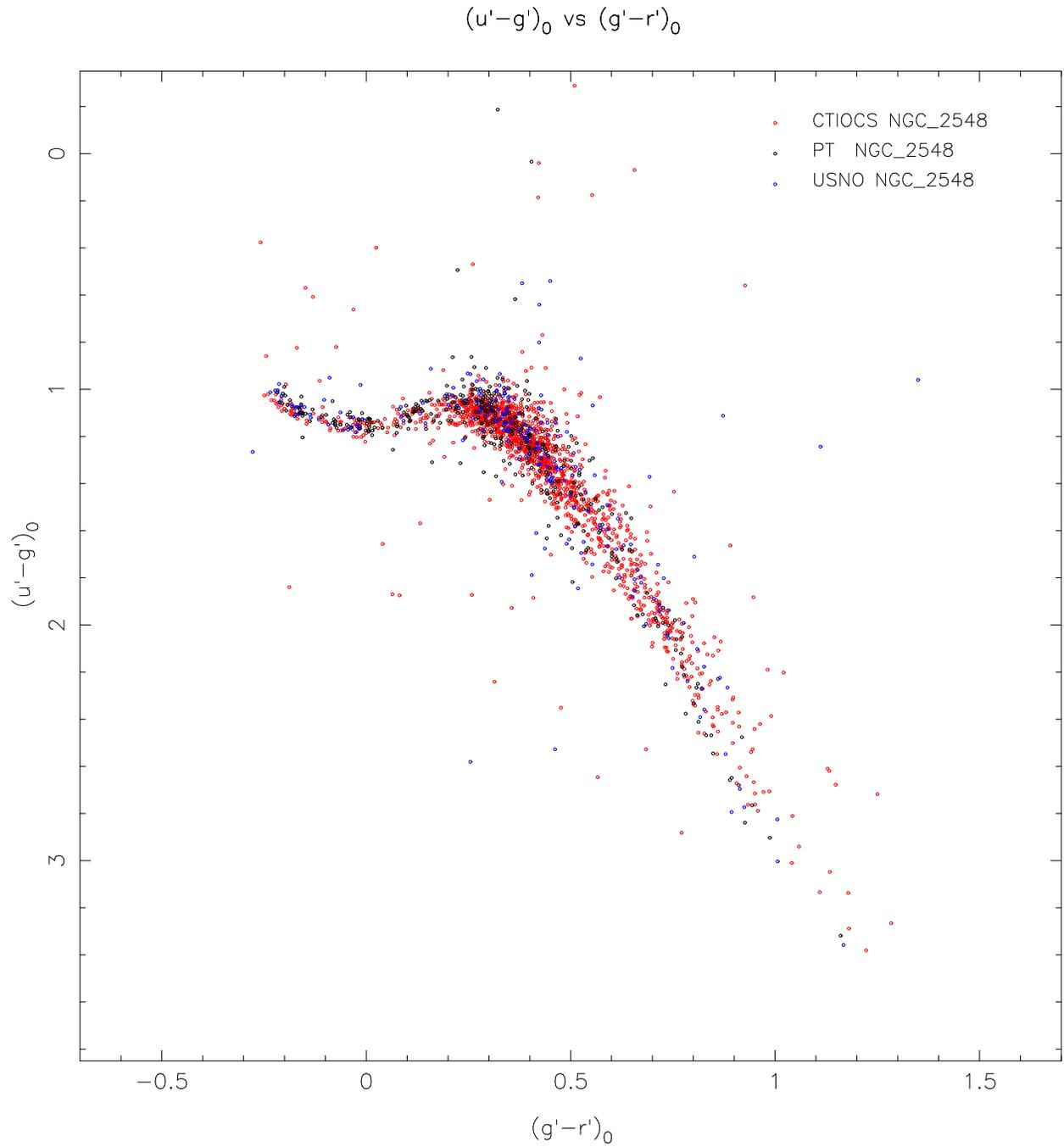}
\caption{A comparison of the de-reddened $u'-g'$ vs. $g'-r'$ color-color diagrams for 
NGC~2548 obtained from the CTIO-CS (red dots), the PT (black dots), and the USNO~1.0m
(blue dots).  
\label{m2548ug}}
\end{figure}

To   further     illustrate these  results,    Figures~\ref{m2548} and
~\ref{m2548ug} show the  results for each  telescope over-plotted in a
color-magnitude   diagram (Figure ~\ref{m2548})  and  in a color-color
diagram (Figure   ~\ref{m2548ug}).    The      data for   the     CTIO
Curtis-Schmidt, the  PT and the USNO 1.0~m  telescopes are  plotted in
red,  black, and  blue, respectively.  These  figures demonstrate that
the stars  observed by each of the  three telescopes fall in  the same
locus and follow the same  trends in both  the color-magnitude and the
color-color diagrams.

\placefigure{m2548}

\placefigure{m2548ug}

These   results verify that    we  can  tie observations   from  other
telescopes to the USNO 1.0~m $u'g'r'i'z'$ system with a high degree of
accuracy (typically $<$2--3\% systematics).

\subsection{The Photometric Properties of NGC~2548}

We compare our observations of NGC~2548 to theoretical SDSS isochrones
and   metallicity  curves     from   \citet{Girardi03}    (see    also
\citealt{Girardi00,   Girardi01, Girardi02}).  The  input physics  for
these models are based upon a magnetic  hydrodynamic equation of state
at temperatures T$ < 10^{7}$ and a fully-ionized gas equation of state
at  higher temperatures; electron  screening   is incorporated in  the
reaction rates.   The  theoretical evolutionary tracks  were converted
into  the  SDSS photometric  system  using   the SDSS 2.5~m  telescope
$ugriz$       filter        response           functions\footnote{{\tt
http://archive.stsci.edu/sdss/documents/response.dat}}     and     the
no-overshoot  ATLAS9  synthetic   atmospheres of   \citet{castelli97}.
Updated versions  of these isochrones  for a  variety of metallicities
are              available                    at:                 {\tt
http://pleiadi.pd.astro.it/$\sim$lgirardi/isoc\_photsys.00/isoc\_sloan/}.

Note that the \citet{Girardi03} SDSS isochrones were created using the
SDSS 2.5~m telescope  $ugriz$ filter  system, which differs   slightly
from the USNO~1.0~m   $u'g'r'i'z'$  filter system  (see,  for  example
\citealt{abazajian03}); furthermore, \citet{Girardi03} assume that the
SDSS 2.5~m $ugriz$   system is a    perfect $AB$ system. In  order  to
compare  the  isochrones to our   data   we first  had to  adjust  the
isochones for the known  deviation of the SDSS  2.5~m telescope from a
true  $AB$ system.  This was  done  using the following equations  (D.
Eisenstein, private communication):
\begin{eqnarray}
u(AB,2.5~m) & = & u(2.5~m) - 0.040\;, \\
g(AB,2.5~m) & = & g(2.5~m) - 0.009\;, \\
r(AB,2.5~m) & = & r(2.5~m)\;, \\
i(AB,2.5~m) & = & i(2.5~m) + 0.017 \;, \\
z(AB,2.5~m) & = & z(2.5~m) + 0.035\ .
\end{eqnarray}
(The values of the $AB$ offsets in these equations are preliminary and
future refinement at the $\pm$ 0.01--0.02~mag level are possible.)

Next, the isochrones were converted from the SDSS 2.5 m $ugriz$ system
into the USNO 1.0~m $u'g'r'i'z'$ system by making use of the following
relations \citep{Tucker03} :

\begin{eqnarray}
u(2.5~m) & = & u'\;, \\
g(2.5~m) & = & g' + 0.060((g'-r')-0.53)\;, \\
r(2.5~m) & = & r' + 0.035((r'-i)'-0.21)\;, \\
i(2.5~m) & = & i' + 0.041((r'-i')-0.21)\;, \\
z(2.5~m) & = & z' - 0.030((i'-z')-0.09)\ .
\end{eqnarray}

The stars  were dereddened using  a value  of $E(B-V)=0.31$  from  the
\citet{Dias02} open  cluster catalog and  the following equations from
\citet{sto02}:
\begin{eqnarray}
   u' & = & u'_{red} - 5.155 \times E(B-V)\;, \\
   g' & = & g'_{red} - 3.793 \times E(B-V)\;, \\
   i' & = & i'_{red} - 2.086 \times E(B-V)\;, \\
   r' & = & r'_{red} - 2.751 \times E(B-V)\;, \\
   z' & = & z'_{red} - 1.479 \times E(B-V)\ .
\end{eqnarray}

The  stars  were then compared   to the results of  the \citet{Dias01}
proper  motion   and membership  study of  this   cluster.  Stars with
membership  probabilities  greater  than 50\%   were considered  to be
members;  stars   with membership probabilities    less than 50\% were
considered non-members.  Further, the stars were  matched to a list of
known red  giants  and spectroscopic  binaries  listed in {\tt webda}.
This matchup information helped in fitting the isochrones to the data.

\begin{figure}
\begin{picture}(600,500)
\put (  0,0){\includegraphics[angle=0,scale=0.33]{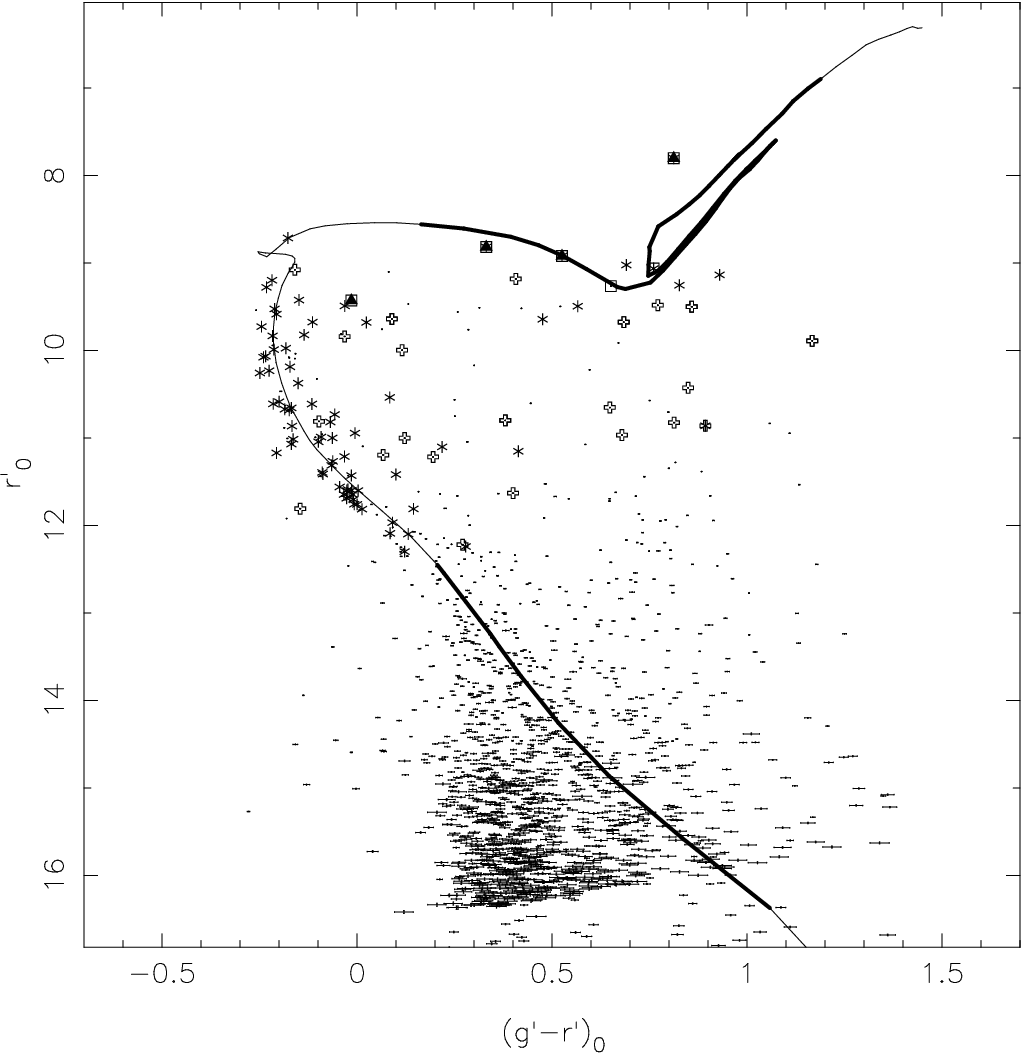}}
\put (250,0){\includegraphics[angle=0,scale=0.33]{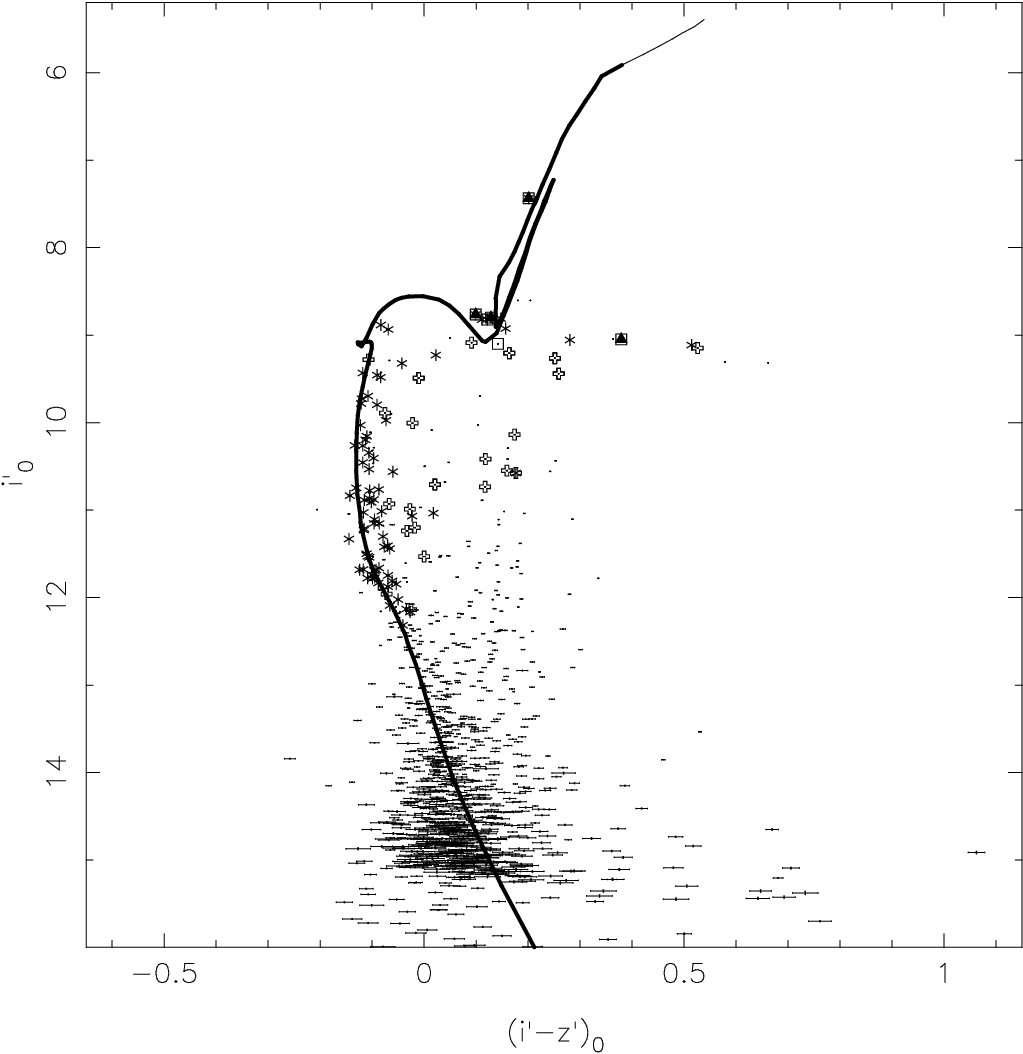}}
\put (0,180){\includegraphics[angle=0,scale=0.33]{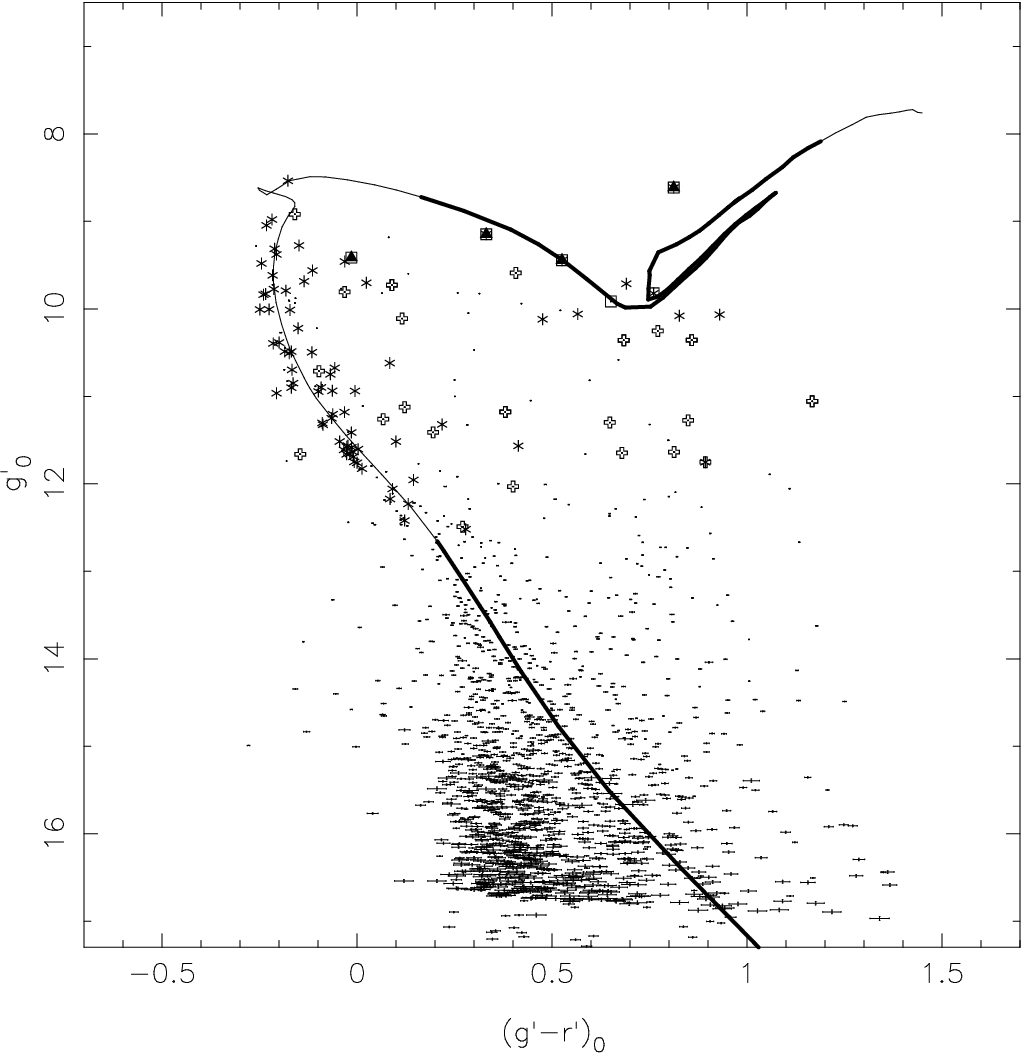}}
\put (250,180){\includegraphics[angle=0,scale=0.33]{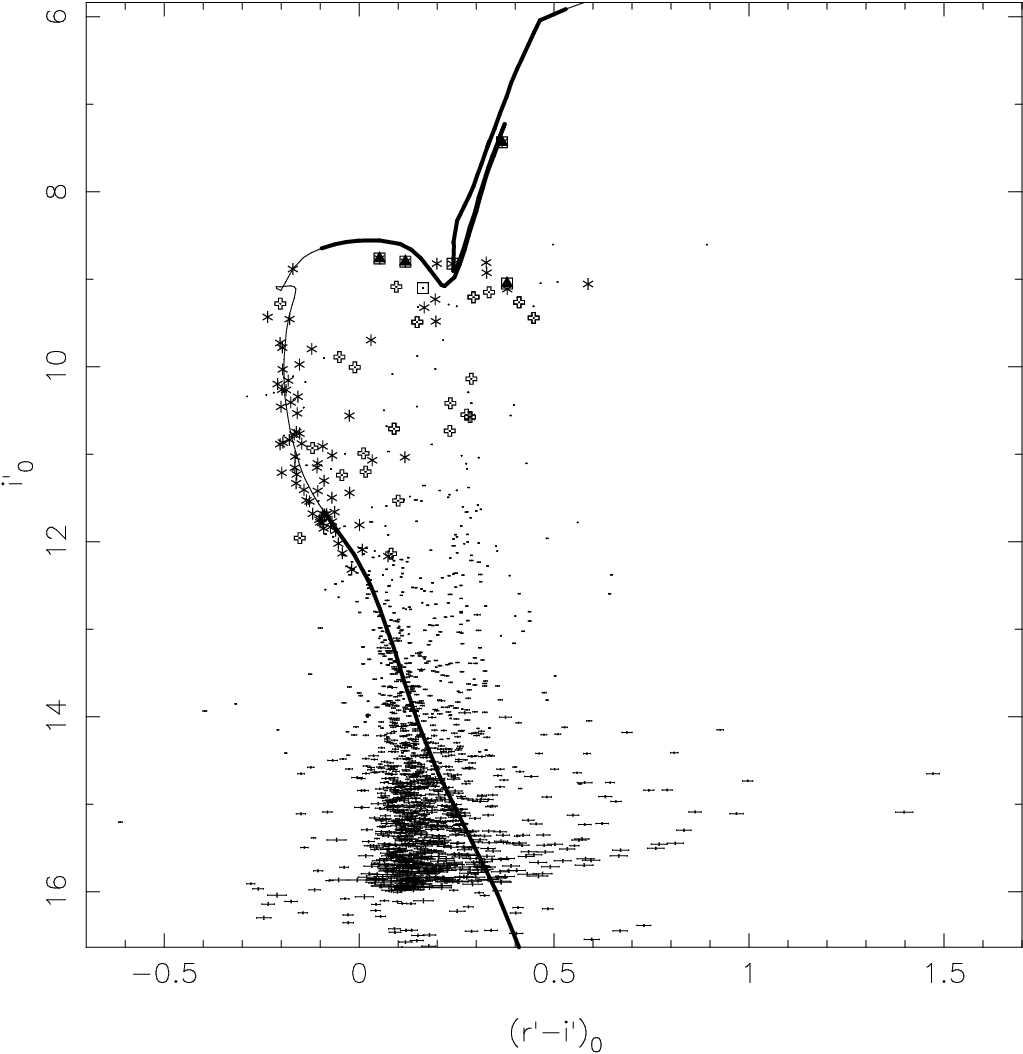}}
\put (0,360){\includegraphics[angle=0,scale=0.33]{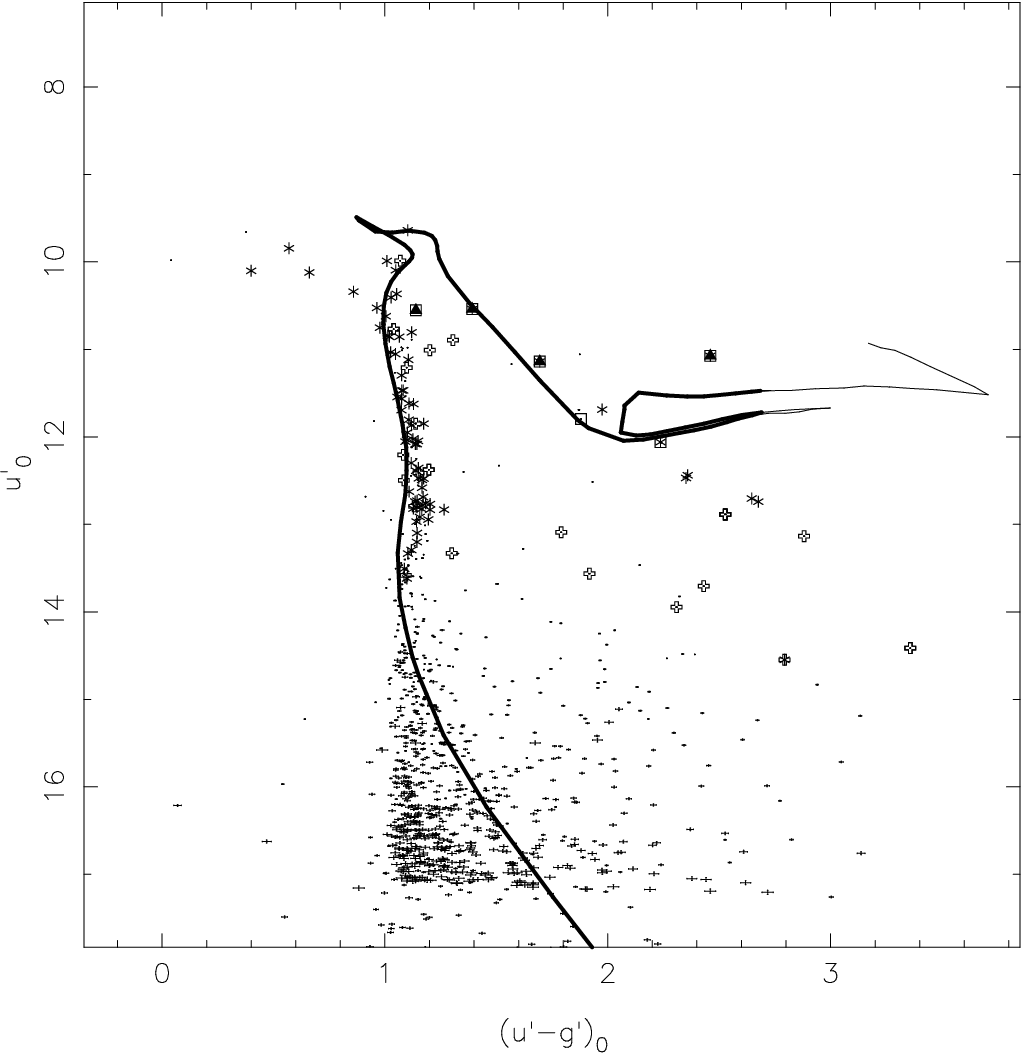}}
\put (250,360){\includegraphics[angle=0,scale=0.33]{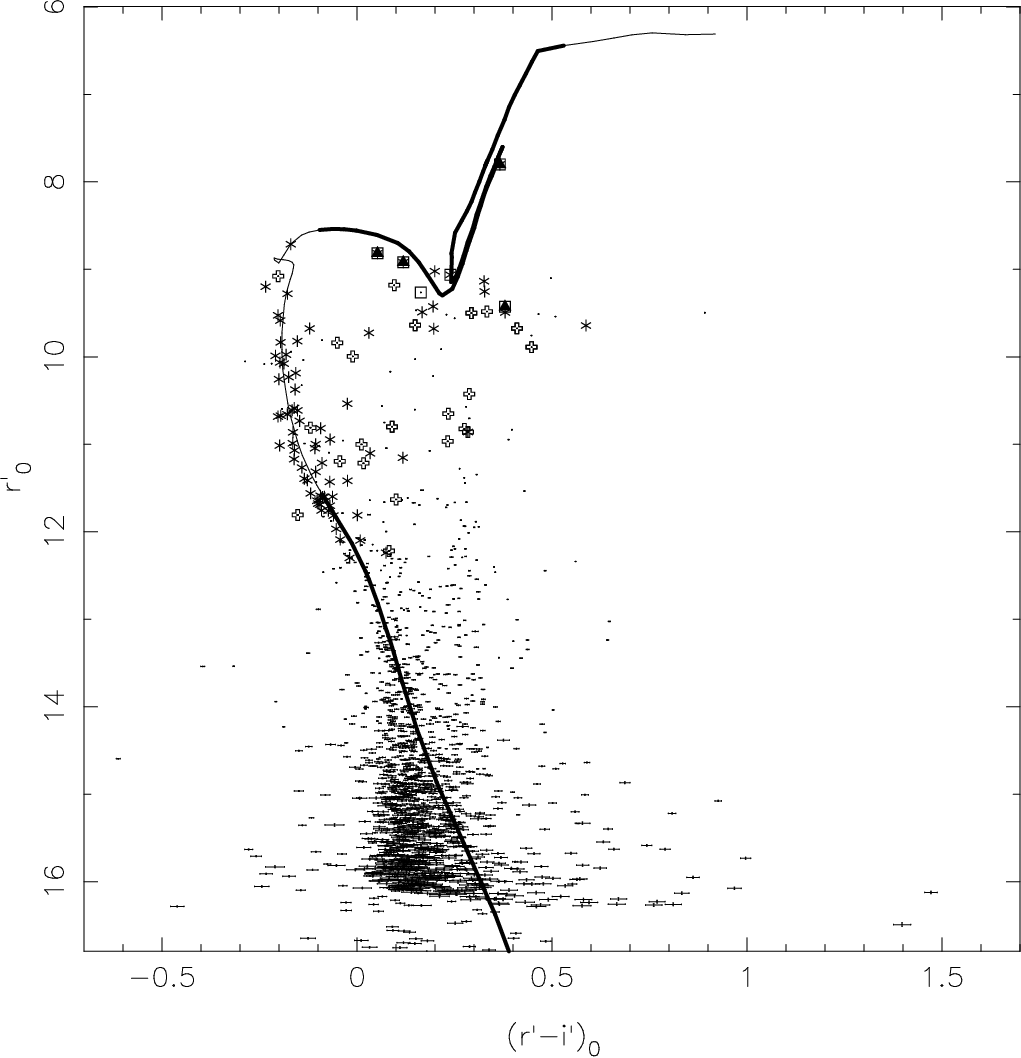}}
\put (135,510) { (a)}
\put (135,330) { (b)}
\put (135,150) { (c)}
\put (385,510) { (d)}
\put (385,330) { (e)}
\put (385,150) { (f)}
\end{picture}
\caption{The de-reddened color-magnitude diagrams for NGC~2548 using 
the  combined data from  all three telescopes.   The solid line is the
0.40~Gyr and $[Z/Z_\odot]=0.0$ isochrone from \citet{Girardi01}. Stars
judged to  be  proper motion members,  proper  motion non-members, red
giants, and spectroscopic binaries  are plotted as asterisks, crosses,
boxes and triangles, respectively.
\label{mosaicCM}}
\end{figure}
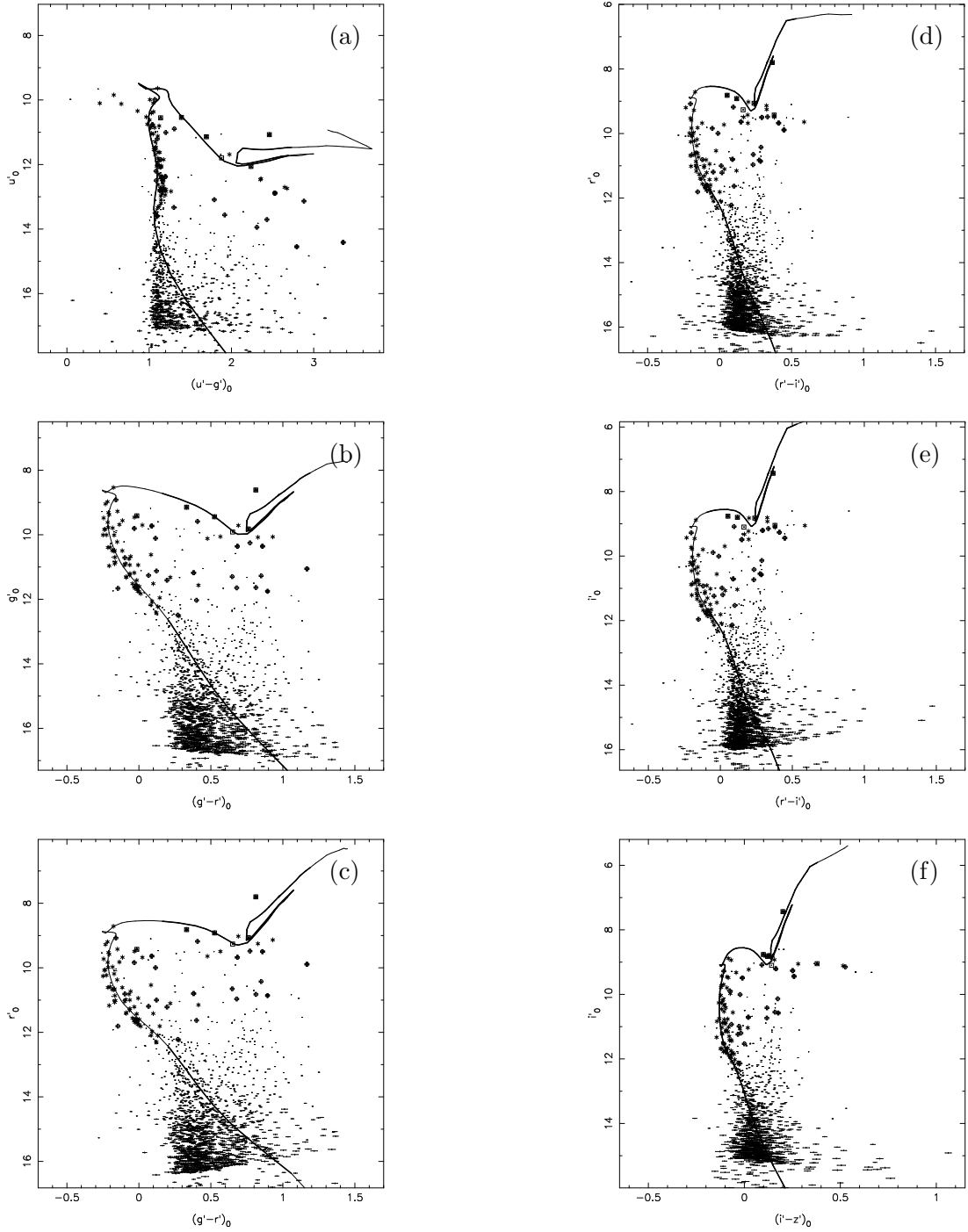

Figures  \ref{mosaicCM}(a-f)   show  the  full set  of color-magnitude
diagrams available    in the $u'g'r'i'z'$   filter   system.  The $g'$
vs. $g'-r'$ (Figure \ref{mosaicCM}b) and the $r'$ vs.  $g'-r'$ (Figure
\ref{mosaicCM}c) color-magnitude diagrams are most similar to the more
recognizable $V$ vs.  $B-V$ color-magnitude diagrams.  (See Table~7 of
\citealt{Smith02}  for   the  transformation  equations   between the
$u'g'r'i'z'$  and the  Johnson   $UBVR_{\rm c}I_{\rm   c}$ photometric
systems.)  All the  color-magnitude diagrams have  a well defined main
sequence and main sequence turn off.

\placefigure{mosaicCM}

\begin{figure}
\begin{picture}(150,150)
\put (  0,30){\includegraphics[angle=0,scale=0.5]{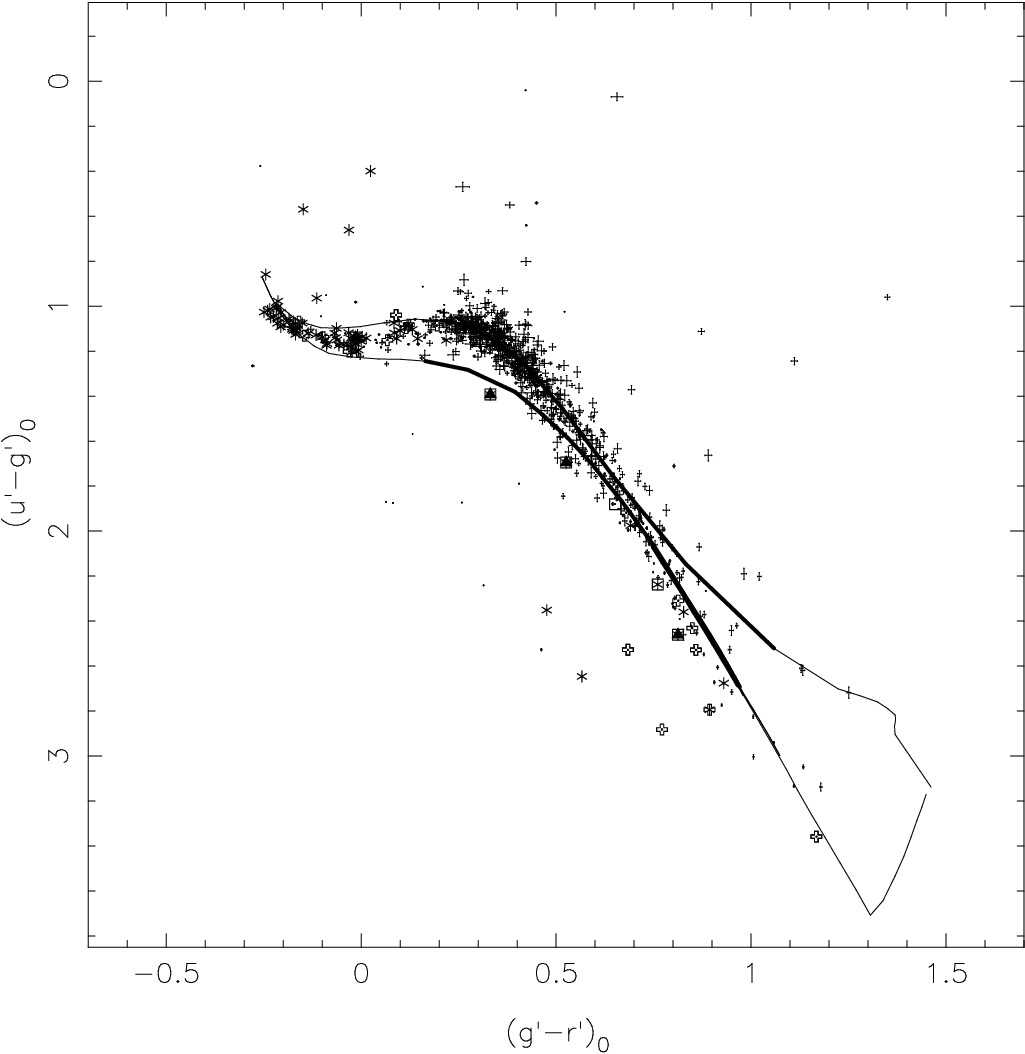}}
\put (270,10){\includegraphics[angle=0,scale=0.25]{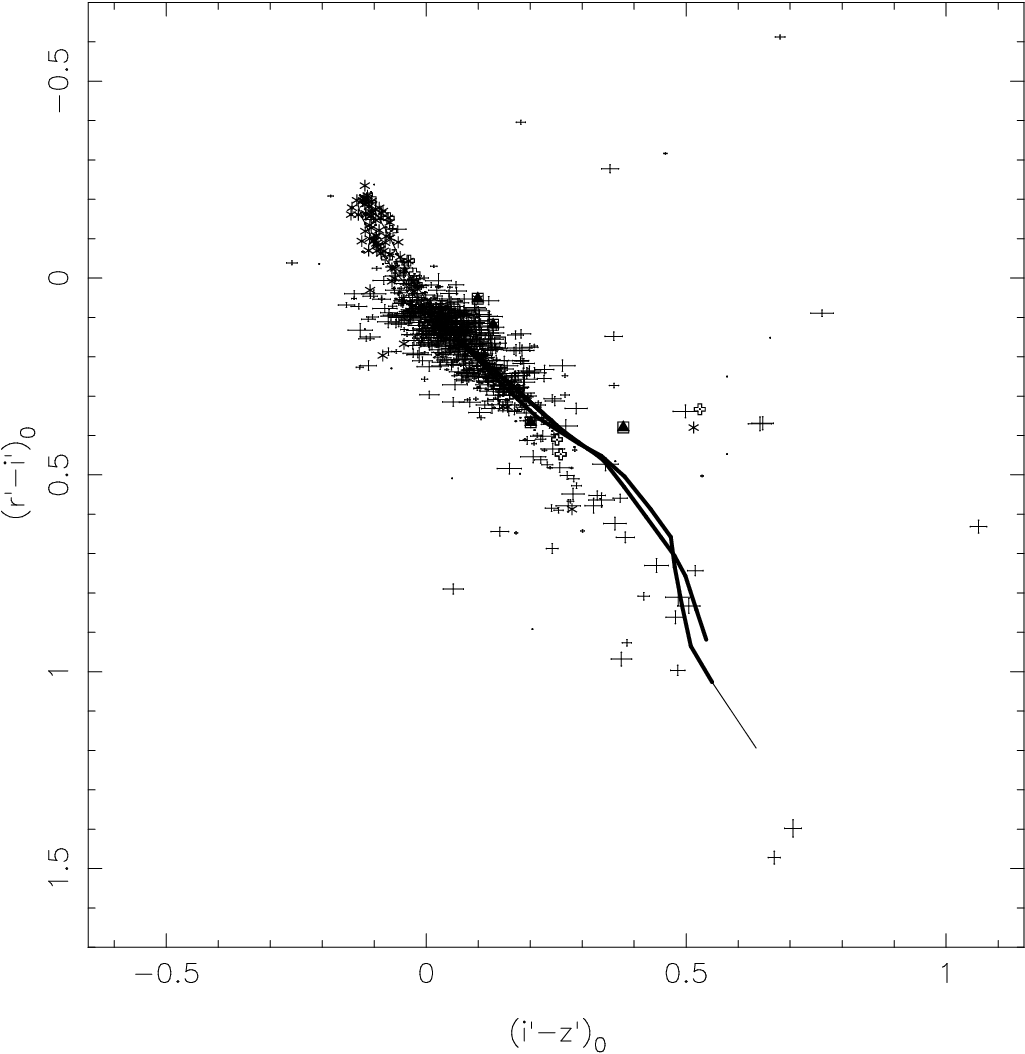}}
\put (270,175){\includegraphics[angle=0,scale=0.25]{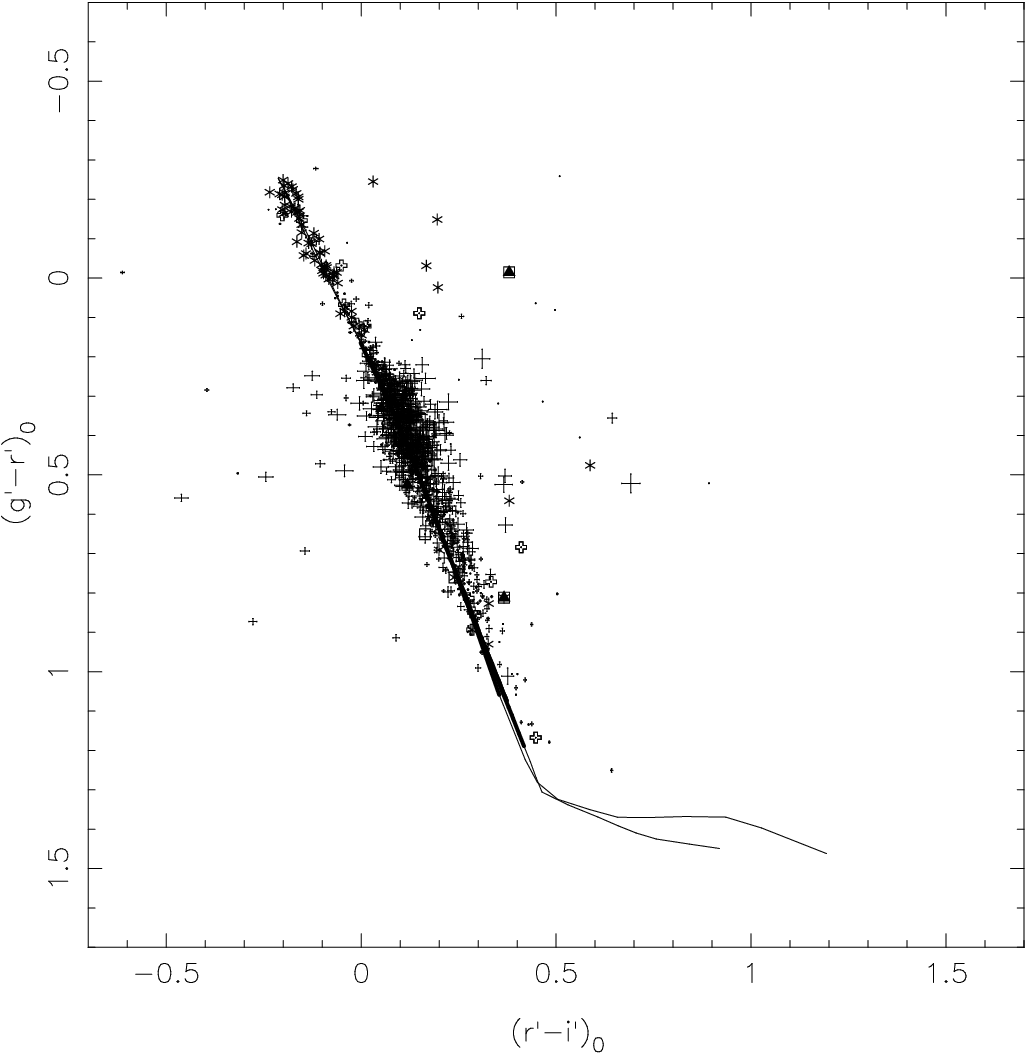}}
\put (120,0){(a)}
\put (330,0){(c)}
\put (330,160){(b)}
\end{picture}
\caption{The de-reddened color-color diagrams for NGC~2548 using the 
combined data from all three telescopes.  To reduce contamination from
field  stars, only stars with   estimated photon noise  errors in  the
colors less than 0.05~mag  are included (typically stars fainter  than
$g'\approx 15$).  The solid line is the 0.40~Gyr and $[Z/Z_\odot]=0.0$
isochrone from  \citet{Girardi01}.   Stars judged to  be proper motion
members,  proper  motion  non-members,  red giants,  and spectroscopic
binaries are  plotted as   asterisks, crosses, boxes   and  triangles,
respectively.
\label{mosaicCC}}
\end{figure}
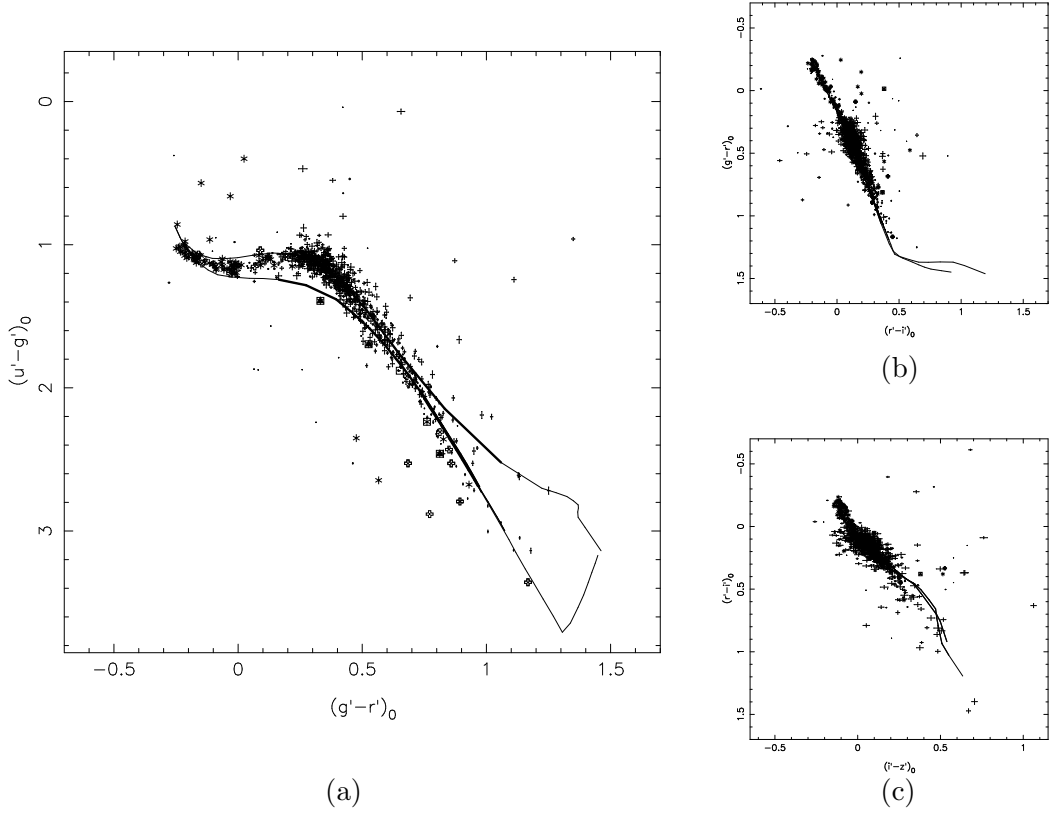

Figures \ref{mosaicCC}(a-c) show the three color-color diagrams of the
$u'g'r'i'z'$    filter   system.    The    $u'-g'$    vs.      $g'-r'$
(Figure~\ref{mosaicCC}a)  color-color  diagram is  most similar to the
more  common $U-B$  vs.   $B-V$ color-color  diagram.  In the  $u'-g'$
vs. $g'-r'$ color-color  diagram the main sequence  (top line) and the
red giant branch  (bottom line) are clearly distinguishable;  however,
in  the $g'-r'$ vs.   $r'-i'$ (Figure~\ref{mosaicCC}b) and $r'-i'$ vs.
$i'-z'$ (Figure
\ref{mosaicCC}c) color-color diagrams the   main sequence and the  red
giant branch are degenerate.

\placefigure{mosaicCC}

We initially adopted the  \citet{Dias02} values for distance, age, and
metallicity of NGC~2548 (769~pc, 0.36~Gyr, +0.08, respectively) as the
starting point for fitting  the theoretical models.  We then  adjusted
these values to find  the best fit  by  eye.  This  was done by  first
adjusting the distance  to fit the  main sequence, and then  adjusting
the age and metallicity to best fit the main sequence turn off and the
red giant branch.  Based upon the  \citet{Girardi03} isochones we find
that the cluster data  are consistent with  a distance in the range of
575~pc to 625~pc, an age in the  range of 0.37~Gyr  to 0.42~Gyr, and a
metallicity  in the range of $[Z/Z_\odot]=-0.1$  to  +0.1; however, we
find that a distance  of 700~pc, an age  of 0.40~Gyr and a metallicity
of $[Z/Z_\odot]=0.0$ best fit our data.  We do note that while some of
the  diagrams indicates a  slightly older age  (mainly the $u'-g'$ vs.
$g'-r'$ color-color  diagram), the majority of  the diagrams  are best
fit by  the 0.40~Gyr  isochrone.    The slight discrepancies   in  the
best-fit parameters among these diagrams  are within the uncertainties
of the   isochrones and  the   $AB$ offsets.    Our finding  for   the
metallicity,  age, and distance for NGC~2548   are consistent with the
\citet{Dias02} values.  We find an  age that is 0.04~Gyr (10\%)  older
and a distance that is 70~pc (10\%) closer than the values listed in
\citet{Dias02}.  Each of    the   color-magnitude diagrams  and    the
color-color diagram  (Figures \ref{mosaicCM}  and \ref{mosaicCC}) have
an  over-plotted  isochrone  with these best   fit values.    The bold
portion of the isochrone represent the region in  which the $ugriz$ to
$u'g'r'i'z'$ transformations are best characterized.  Overall, we find
the
\citet{Girardi03} isochrones  fit our data in  a consistent manner and
can be used  to extract meaningful  cluster information (e.g., age and
distance).  This holds true even bluewards of region where the $ugriz$
to $u'g'r'i'z'$ transformations are best characterized.

\section{ Summary and Future Work}

In this first paper of our series, we show that we can effectively tie
$u'g'r'i'z'$  observations from  other   telescopes to  those  by  the
USNO~1.0~m telescope and CCD detector --- the configuration which {\em
defined} $u'g'r'i'z'$ photometric system ---   with a high degree   of
accuracy    (typically  $<$2--3\%  systematics).   This    lays a firm
foundation  for  the rest of our   open cluster project, which derives
data primarily from the CTIO Curtis-Schmidt  telescope.  Also, we have
shown  that  we  can effectively   fit  isochrones and  extract useful
cluster information   using the $u'g'r'i'z'$  filter system.   We used
these isochrones to determine  the age, distance, and  metallicity for
NGC~2548.  Our  metallicity   value ($[Z/Z_\odot]=0.0$)  is consistent
with the currently accepted  value listed  in  the {\tt webda}  online
database and in the \citep{Dias02}  open  cluster catalog, as are  our
age and and  distance  values.  We  find  a distance of  700~pc  (10\%
closer than   the value reported in  \citealt{Dias02})   and an age of
0.40~Gyr (10\% older than the \citealt{Dias02} value.)

Our entire  sample  of observations  consists of   more  than 100 open
clusters.  Our immediate goal is to  publish a sample of open clusters
covering a range  of ages and metallicities  to help better  determine
the characteristics  of the $u'g'r'i'z'$  filter  system.  To date, we
have  built and tested the cluster  data matching scripts and adjusted
isochrone  fits.  We  will pursue  the reduction and   analyses of the
remaining  clusters for which  we   have data.    As  the  SDSS  2.5~m
telescope  sweeps   across  open  clusters, we'll    incorporate those
observations into our survey.  Further, as time and conditions permit,
the SDSS PT may be used for a survey of northern  clusters in much the
same  way the CTIO Schmidt telescope  was used in this current effort.
It is relatively small (0.5~m)  but has a  large field of view, making
it an ideal instrument for the  initial observations of the larger and
brighter clusters.   Follow-up work  for fainter clusters  and cluster
members will be pursued at available observatories as time permits.

Repeat observations for some of the clusters will be pursued to expand
the  investigations to  find and  analyze  variable  stars (especially
RR-Lyr type  stars to improve   distance estimates).  Deeper exposures
for  some clusters will be  obtained to improve  the model fits to the
lower main sequence and   reduce the scatter that   is present in  our
reconnaissance images.

\acknowledgments

The authors would  like to thank  the tremendous support  given by the
mountain  staff at CTIO,  especially  by  Edgardo Cosgrove and  Arturo
Gomez.   We would  also  like to  thank  Jeff Pier  and Jeff  Munn for
providing access and support  for our observations  with the  US Naval
Observatory  1~m telescope  at   Flagstaff Station, Arizona,  and Alan
Uomoto of Johns  Hopkins  University for providing  a  working monitor
telescope (in the form of the PT) at Apache Point Observatory.  We all
gratefully  acknowledge the support of  Leo Girardi for his models and
suggestions to improve the fits to the data.  We also extend thanks to
Daniel Eisenstein, Eva Grebel, Christy Tremonti, and the other members
of the  SDSS Stars   and  Calibration Working  Groups for  suggestions
regarding the  correct  values of  the  $AB$  offsets to   use in  our
isochrone fits.  We further acknowledge  the {\tt webda} open  cluster
data  compilation which made  accessing  all  available data  for this
clusters an easy task.

CJR   acknowledges  support  from   the   Fermi National   Accelerator
Laboratory Internship  for Undergraduate Physics Majors program during
the summer of 2002.

JAS acknowledges  support from an  American Astronomical Society Small
Research  Grant  and the  National   Science Foundation  through grant
AST-0097356, which allowed the observations to  be obtained.  JAS also
acknowledges the University of  Michigan, Department of  Astronomy for
access to the Curtis-Schmidt telescope while a  research fellow at the
University.   JAS also acknowledges   the  proof reading support   and
English skills   of   Jeanne Odermann,  whose  efforts   substantially
improved the readability of the text.

This research has made use of the NASA Astrophysics Data System and of
the  Guide Star Catalog  2.2.  The Guide  Star Catalog was produced at
the Space Telescope Science Institute under U.S.  Government grant.

Funding for the  Sloan Digital Sky Survey  (SDSS) has been provided by
the  Alfred P. Sloan  Foundation,  the Participating Institutions, the
National Aeronautics and  Space  Administration, the National  Science
Foundation,  the     U.S.    Department of   Energy,     the  Japanese
Monbukagakusho,  and  the Max  Planck Society.  The  SDSS  Web site is
{\tt http://www.sdss.org/}.

The SDSS is managed by the Astrophysical Research Consortium (ARC) for
the Participating Institutions. The Participating Institutions are The
University of Chicago, Fermilab, the Institute for Advanced Study, the
Japan Participation Group,  The  Johns Hopkins University,  Los Alamos
National  Laboratory, the  Max-Planck-Institute for Astronomy  (MPIA),
the  Max-Planck-Institute  for  Astrophysics  (MPA),  New Mexico State
University, University of Pittsburgh, Princeton University, the United
States Naval Observatory, and the University of Washington.

%\newpage

%References

\clearpage

\clearpage

\clearpage

\clearpage

\clearpage

\clearpage

\clearpage

\clearpage

\clearpage

\clearpage

\begin{deluxetable}{lcccccc}
\rotate
\tabletypesize{\footnotesize}
\tablecaption{Telescope Parameters. \label{systems}}
\tablewidth{0pt}
\tablenum{1}
\tablehead{
 \colhead{Telescope} &
 \colhead{Aperture}  &
 \colhead{Focal Length} &
 \colhead{Chip} &
 \colhead{\# of Amplifiers}  &
 \colhead{Gain (epadu)}   &
 \colhead{Read Noise (e$^{-}$)}
}
\startdata
USNO 	&  1.0 m   & f/7.3  & TK1024 	& 1 & 7.43$\pm$.41 		 & 6.0  \\
PT\tablenotemark{*} 	&  0.5 m    & f/8.8 &SITe 2048x2048   & 2 & 4.89$\pm$.03, 4.44$\pm$0.02 & 7.43 , 7.86   \\
CTIO-CS 	&0.6/0.9 m & f/3.5  &  TeK2K\#5 & 2 &  2.89 (left) 2.90 (right)  &  4.1 (left) 4.3 (right)   \\
\enddata
\tablenotetext{*}{measurements of PT gain and readnoise from Fukugita, M., Ichikawa, S., Watanabe, M., \& Yasuda, N., 2000, ``Verification of the SDSS Photometric Telecope: I. Electronics,'' (unpublished SDSS internal document).}

\end{deluxetable}

\clearpage

\begin{deluxetable}{lcccccccl}
%\tabletypesize{\large}
%\tabletypesize{\scriptsize}
\tablenum{2}
\tablecaption{Observing Circumstances\label{obs}}
%\tablewidth{0pt}
\tablehead{
  \colhead{YYMMDD} &
  \colhead{MJD} &
  \colhead{Airmass} &
  \multicolumn{5}{c}{Exposure (sec)} &
  \colhead{Comments} \\
  \colhead{} &
  \colhead{} &
  \colhead{} &
  \colhead{ $r'$ } &
  \colhead{ $g'$ } &
  \colhead{ $u'$ } &
  \colhead{ $i'$ } &
  \colhead{ $z'$ } &
  \colhead{}
}
\startdata
\multicolumn{9}{c}{USNO 1.0~m}\\
021105 & 52583 & 1.425   &  10      &  10      &  60      & 10      & 15        & Photometric \\
       &       & 1.408   &  60      &  60      & 360      & 60      & 90        & Photometric \\
       &       & 1.349   &  10      &  10      &  60      & 10      & 15        & Photometric \\
       &       & 1.341   &  60      &  60      & 360      & 60      & 90        & Photometric \\
021106 & 52584 & 1.430   &  10      &  10      &  60      & 10      & 15        & Photometric \\
       &       & 1.405   &  60      &  60      & 360      & 60      & 90        & Photometric \\
       &       & 1.369   &   5      &   5      &  30      &  5      &  8        & Photometric \\
\multicolumn {9}{c}{Photometric Telescope}\\
011202 & 52245 & 1.339   &  15      &  15      &  90      & 15      & 30        & Photometric \\
\multicolumn {9}{c}{CTIO Curtis-Schmidt}\\
000306 & 51609 & 1.125   &   5      &   5      &  30      &  5      &  5        & Photometric \\
       &       & 1.137   &  30      &  30      & 240      & 30      & 30        & Photometric \\

\enddata

\end{deluxetable}

\clearpage

\begin{deluxetable}{lcccccccc}
%\rotate
%\tabletypesize{\footnotesize}
\tabletypesize{\tiny}
\tablecaption{Night Characterization Coefficients\tablenotemark{a}\label{nightChar}}
\tablenum{3}
\tablewidth{0pt}
\tablehead{
  \colhead{MJD} &
  \colhead{Filter} &
  \colhead{Zeropoint (a)} &
  \colhead{Instr. Color (b)} &
  \multicolumn{3}{c}{1st-Order Ext. (k)} &
  \colhead{Std. rms} &
  \colhead{\# Std.} \\
  \colhead{}  &
  \colhead{}  &
  \colhead{}  &
  \colhead{}  &
  \colhead{block 0} &
  \colhead{block 1} &
  \colhead{block 2} &
  \colhead{(mag)}  &
  \colhead{stars}  \\
  \colhead{(1)}  &
  \colhead{(2)}  &
  \colhead{(3)}  &
  \colhead{(4)}  &
  \colhead{(5)}  &
  \colhead{(6)}  &
  \colhead{(7)}  &
  \colhead{(8)}  &
  \colhead{(9)}
}
\startdata
\multicolumn{9}{c}{USNO 1.0~m\tablenotemark{b}}                                                                                                                \\
      &      &                    &                    & 01:47-04:47UT   & 04:47-07:47UT   & 07:47-12:43UT   &       &    \\
52583 & $u'$ & -20.036$\pm$0.062  &   0.000            & 0.434$\pm$0.050 & 0.459$\pm$0.041 & 0.432$\pm$0.037 & 0.044 & 19 \\
52583 & $g'$ & -21.654$\pm$0.014  &   0.000            & 0.155$\pm$0.011 & 0.150$\pm$0.010 & 0.147$\pm$0.008 & 0.011 & 20 \\
52583 & $r'$ & -21.517$\pm$0.010  &   0.000            & 0.094$\pm$0.008 & 0.091$\pm$0.007 & 0.086$\pm$0.006 & 0.007 & 19 \\
52583 & $i'$ & -21.079$\pm$0.012  &   0.000            & 0.039$\pm$0.009 & 0.035$\pm$0.008 & 0.033$\pm$0.007 & 0.008 & 19 \\
52583 & $z'$ & -19.974$\pm$0.020  &   0.000            & 0.057$\pm$0.016 & 0.033$\pm$0.013 & 0.029$\pm$0.011 & 0.014 & 20 \\
      &      &                    &                    &                 &                 &                 &       &    \\
      &      &                    &                    & 01:29-04:29UT   & 04:29-07:29UT   & 07:29-13:10UT   &       &    \\
52584 & $u'$ & -20.200$\pm$0.044  &   0.000            & 0.556$\pm$0.033 & 0.552$\pm$0.030 & 0.550$\pm$0.032 & 0.039 & 18 \\
52584 & $g'$ & -21.686$\pm$0.021  &   0.000            & 0.180$\pm$0.016 & 0.178$\pm$0.014 & 0.179$\pm$0.016 & 0.019 & 18 \\
52584 & $r'$ & -21.499$\pm$0.012  &   0.000            & 0.088$\pm$0.009 & 0.087$\pm$0.008 & 0.078$\pm$0.009 & 0.010 & 17 \\
52584 & $i'$ & -21.068$\pm$0.013  &   0.000            & 0.043$\pm$0.010 & 0.041$\pm$0.009 & 0.036$\pm$0.010 & 0.012 & 17 \\
52584 & $z'$ & -19.943$\pm$0.020  &   0.000            & 0.049$\pm$0.015 & 0.04$\pm$10.013 & 0.026$\pm$0.015 & 0.018 & 18 \\
      &      &                    &                    &                 &                 &                 &       &    \\
\multicolumn{9}{c}{Photometric Telescope\tablenotemark{c}}                                                                \\
      &      &                    &                    & 02:20-11:12UT   &                 &                 &       &    \\    
52245 & $u'$ & -18.959$\pm$0.030  &   0.001            & 0.476$\pm$0.020 & \nodata         & \nodata         & 0.034 & 31 \\
52245 & $g'$ & -20.776$\pm$0.011  &  -0.041            & 0.137$\pm$0.007 & \nodata         & \nodata         & 0.012 & 32 \\
52245 & $r'$ & -20.686$\pm$0.008  &   0.009            & 0.088$\pm$0.005 & \nodata         & \nodata         & 0.009 & 33 \\
52245 & $i'$ & -20.215$\pm$0.011  &   0.010            & 0.043$\pm$0.008 & \nodata         & \nodata         & 0.013 & 34 \\
52245 & $z'$ & -19.266$\pm$0.014  &   0.002            & 0.048$\pm$0.009 & \nodata         & \nodata         & 0.016 & 34 \\
      &      &                    &                    &                 &                 &                 &       &    \\
\multicolumn{9}{c}{CTIO Curtis-Schmidt}                                                                                   \\
      &      &                    &                    & 00:27-09:36UT   & \nodata         & \nodata         &       &    \\
51609 & $u'$ & -18.876$\pm$0.023  &  -0.020$\pm$0.008  & 0.516$\pm$0.014 & \nodata         & \nodata         & 0.020 & 23 \\
51609 & $g'$ & -20.864$\pm$0.012  &   0.013$\pm$0.007  & 0.203$\pm$0.007 & \nodata         & \nodata         & 0.010 & 23 \\
51609 & $r'$ & -20.794$\pm$0.023  &   0.000$\pm$0.032  & 0.116$\pm$0.014 & \nodata         & \nodata         & 0.019 & 23 \\
51609 & $i'$ & -20.364$\pm$0.027  &  -0.025$\pm$0.049  & 0.077$\pm$0.015 & \nodata         & \nodata         & 0.022 & 24 \\
51609 & $z'$ & -19.387$\pm$0.017  &   0.029$\pm$0.061  & 0.004$\pm$0.001 & \nodata         & \nodata         & 0.026 & 23 \\
\enddata

\tablenotetext{a} {The values for the second order extinction term ($c$) 
coefficients were set  to $-2.1 \times 10^{-2}$, $-1.6\times 10^{-2}$,
$-4.0 \times 10^{-3}$, $6.0 \times 10^{-3}$  and $3.0 \times 10^{-3}$,
for  the $u'$,$g'$,$r'$,$i'$ and  $z'$  respectively. These values are
those determined in \citet{Smith02}.}

\tablenotetext{b} {The USNO 1.0~m telescope, its Tektronix TK1024 CCD, 
and  its $u'g'r'i'z'$  filter   set  {\em define\/} the   $u'g'r'i'z'$
photometric system; hence, the instrumental   color term ($b$)  values
are set to zero by definition.}

\tablenotetext{c} {We have set the instrumental color term ($b$) 
coefficients for the PT to their site averages rather than solving for
them based upon a single night's data.}

\end{deluxetable}

\clearpage

\begin{deluxetable}{lccccccccccccccccccl}
\rotate
\tablenum{4}
\tabletypesize{\tiny}
%\tabletypesize{\scriptsize}
\tablecaption{Example of Available Data for the USNO 1.0~m  Telescope\label{dataUSNO}}
\tablewidth{0pt}
\tablehead{
  \colhead{Star ID} &
  \colhead{WEBDA ID} &
  \colhead{RA} &
  \colhead{DEC} &
  \multicolumn{5}{c}{Magnitude} &	
  \multicolumn{5}{c}{Magnitude Error} &
  \multicolumn{5}{c}{Saturation Flag} &
  \colhead{Membership } \\	
  \colhead{} &
 \colhead{} &	 
  \colhead{} &
  \colhead{} &
  \colhead{ $u'$ } &
  \colhead{ $g'$ } &
  \colhead{ $r'$ } &
  \colhead{ $i'$ } &
  \colhead{ $z'$ } &
 \colhead{ $u'$ } &
  \colhead{ $g'$ } &
  \colhead{ $r'$ } &
  \colhead{ $i'$ } &
  \colhead{ $z'$ } &
 \colhead{ $u'$ } &
  \colhead{ $g'$ } &
  \colhead{ $r'$ } &
  \colhead{ $i'$ } &
  \colhead{ $z'$ } &
  \colhead{Probability}\\
 \colhead{1} &
 \colhead{2} &
 \colhead{3} &
 \colhead{4} &
 \colhead{5} &
 \colhead{6} &
 \colhead{7} &
 \colhead{8} & 
 \colhead{9} &
 \colhead{10} &
 \colhead{11} &
 \colhead{12} &
 \colhead{13} &
 \colhead{14} &
 \colhead{15} &
 \colhead{16} &
 \colhead{17} &
 \colhead{18} &
 \colhead{19} &
 \colhead{20} 
}
\startdata
3 & 1260 & +08:13:40.3 & -05:46:25 & 10.533 & 9.142 & 8.813 & 8.762 & 8.656 & 0.000 & 0.000 & 0.000 & 0.000 & 0.001 & 0 & 0 & 0 & 0 & 0 & 97 \\
4 & 1296 & +08:13:44.7 & -05:48:01 & 11.132 & 9.439 & 8.916 & 8.799 & 8.663 & 0.001 & 0.001 & 0.000 & 0.000 & 0.001 & 0 & 0 & 0 & 0 & 0 & 96 \\
7 & 1313 & +08:13:46.6 & -05:44:52 & 9.981 & 8.913 & 9.075 & 9.279 & 9.379 & 0.000 & 0.000 & 0.000 & 0.000 & 0.000 & 0 & 0 & 0 & 0 & 0 & 5 \\
11 & 1289 & +08:13:44.1 & -05:48:49 & 9.982 & 8.975 & 9.195 & 9.431 & 9.535 & 0.001 & 0.000 & 0.000 & 0.000 & 0.001 & 0 & 0 & 0 & 0 & 0 & 93 \\
12 & 870 & +08:12:51.2 & -05:50:51 & 12.433 & 10.074 & 9.250 & 8.925 & 8.761 & 0.001 & 0.001 & 0.001 & 0.001 & 0.001 & 0 & 0 & 0 & 0 & 0 & 87 \\
13 & 1218 & +08:13:35.3 & -05:53:02 & 11.789 & 9.910 & 9.262 & 9.100 & 8.951 & 0.001 & 0.001 & 0.001 & 0.001 & 0.001 & 0 & 0 & 0 & 0 & 0 & -1 \\
25 & 978 & +08:13:05.3 & -05:45:01 & 10.361 & 9.309 & 9.523 & 9.727 & 9.839 & 0.000 & 0.001 & 0.001 & 0.000 & 0.000 & 0 & 0 & 0 & 0 & 0 & 96 \\
28 & 1169 & +08:13:28.6 & -05:48:15 & 10.401 & 9.373 & 9.582 & 9.781 & 9.894 & 0.000 & 0.001 & 0.001 & 0.000 & 0.000 & 0 & 0 & 0 & 0 & 0 & 96 \\
37 & 1253 & +08:13:39.5 & -05:47:15 & 10.799 & 9.680 & 9.810 & 9.973 & 10.039 & 0.000 & 0.001 & 0.001 & 0.000 & 0.001 & 0 & 0 & 0 & 0 & 0 & 95 \\
39 & 1117 & +08:13:23.1 & -05:45:23 & 10.616 & 9.612 & 9.831 & 10.029 & 10.144 & 0.000 & 0.001 & 0.001 & 0.000 & 0.001 & 0 & 0 & 0 & 0 & 0 & 97 \\
41 & 1434 & +08:13:59.6 & -05:53:22 & 14.413 & 11.051 & 9.886 & 9.440 & 9.168 & 0.003 & 0.000 & 0.001 & 0.000 & 0.001 & 0 & 0 & 0 & 0 & 0 & 41 \\
44 & 975 & +08:13:04.9 & -05:53:05 & 10.851 & 9.787 & 9.972 & 10.156 & 10.258 & 0.000 & 0.001 & 0.001 & 0.000 & 0.001 & 0 & 0 & 0 & 0 & 0 & 89 \\
45 & 1338 & +08:13:48.9 & -05:44:23 & 10.746 & 9.775 & 9.977 & 10.197 & 10.302 & 0.000 & 0.001 & 0.001 & 0.000 & 0.001 & 0 & 0 & 0 & 0 & 0 & 96 \\
49 & -1 & +08:13:05.1 & -05:45:11 & -100.000 & -100.000 & 10.051 & 10.339 & -100.000 & -100.000 & -100.000 & 0.001 & 0.001 & -100.000 & 1 & 1 & 0 & 0 & 1 & -1 \\
50 & 1147 & +08:13:26.5 & -05:49:54 & 10.836 & 9.824 & 10.053 & 10.261 & 10.386 & 0.000 & 0.001 & 0.001 & 0.000 & 0.001 & 0 & 0 & 0 & 0 & 0 & 95 \\
51 & -1 & +08:13:43.1 & -05:45:53 & 10.970 & 9.897 & 10.075 & 10.295 & 10.394 & 0.000 & 0.000 & 0.000 & 0.000 & 0.001 & 0 & 0 & 0 & 0 & 0 & -1 \\
53 & -1 & +08:13:42.9 & -05:46:00 & 10.983 & 9.904 & 10.081 & 10.320 & 10.413 & 0.000 & 0.000 & 0.000 & 0.000 & 0.001 & 0 & 0 & 0 & 0 & 0 & -1 \\
61 & 1306 & +08:13:45.9 & -05:46:02 & 11.260 & 10.222 & 10.313 & 10.465 & 10.557 & 0.001 & 0.001 & 0.001 & 0.000 & 0.001 & 0 & 0 & 0 & 0 & 0 & -1 \\
62 & 1029 & +08:13:12.1 & -05:46:42 & 11.292 & 10.217 & 10.371 & 10.532 & 10.630 & 0.001 & 0.000 & 0.000 & 0.000 & 0.001 & 0 & 0 & 0 & 0 & 0 & 96 \\
65 & 848 & +08:12:48.5 & -05:53:45 & 11.688 & 10.614 & 10.533 & 10.560 & 10.612 & 0.001 & 0.000 & 0.000 & 0.000 & 0.001 & 0 & 0 & 0 & 0 & 0 & 52 \\
71 & 1362 & +08:13:52.0 & -05:54:20 & 11.623 & 10.491 & 10.609 & 10.763 & 10.838 & 0.001 & 0.000 & 0.000 & 0.001 & 0.001 & 0 & 0 & 0 & 0 & 0 & 91 \\
77 & 1406 & +08:13:57.2 & -05:48:14 & 11.615 & 10.504 & 10.680 & 10.884 & 10.986 & 0.001 & 0.000 & 0.000 & 0.001 & 0.001 & 0 & 0 & 0 & 0 & 0 & 94 \\
79 & 1424 & +08:13:58.9 & -05:50:40 & 11.848 & 10.670 & 10.730 & 10.878 & 10.962 & 0.001 & 0.000 & 0.000 & 0.001 & 0.001 & 0 & 0 & 0 & 0 & 0 & 96 \\
81 & 1454 & +08:14:02.9 & -05:50:26 & 12.374 & 11.173 & 10.796 & 10.708 & 10.680 & 0.001 & 0.000 & 0.000 & 0.001 & 0.001 & 0 & 0 & 0 & 0 & 0 & 0 \\
86 & 1230 & +08:13:36.6 & -05:51:22 & 14.546 & 11.748 & 10.857 & 10.575 & 10.391 & 0.003 & 0.001 & 0.000 & 0.000 & 0.001 & 0 & 0 & 0 & 0 & 0 & 57 \\
91 & -1 & +08:13:38.7 & -05:47:17 & 11.813 & 10.863 & 10.955 & 10.993 & 11.191 & 0.001 & 0.000 & 0.000 & 0.001 & 0.001 & 0 & 0 & 0 & 0 & 0 & -1 \\
93 & -1 & +08:13:40.9 & -05:46:56 & 12.044 & 10.892 & 10.986 & 11.153 & 11.226 & 0.001 & 0.000 & 0.000 & 0.001 & 0.001 & 0 & 0 & 0 & 0 & 0 & 87 \\
95 & 924 & +08:12:58.4 & -05:51:06 & 12.029 & 10.931 & 10.997 & 11.105 & 11.192 & 0.001 & 0.000 & 0.000 & 0.001 & 0.001 & 0 & 0 & 0 & 0 & 0 & 95 \\
98 & 1268 & +08:13:41.3 & -05:50:55 & 11.949 & 10.845 & 11.011 & 11.211 & 11.314 & 0.001 & 0.000 & 0.000 & 0.001 & 0.001 & 0 & 0 & 0 & 0 & 0 & 96 \\
108 & 1124 & +08:13:23.6 & -05:50:06 & 12.296 & 11.175 & 11.210 & 11.301 & 11.367 & 0.001 & 0.000 & 0.001 & 0.001 & 0.001 & 0 & 0 & 0 & 0 & 0 & 97 \\
111 & 1339 & +08:13:49.0 & -05:46:25 & 14.485 & 12.097 & 11.276 & 11.015 & 10.848 & 0.003 & 0.001 & 0.001 & 0.001 & 0.001 & 0 & 0 & 0 & 0 & 0 & -1 \\
114 & 2529 & +08:13:35.9 & -05:48:34 & 14.530 & 12.268 & 11.381 & 11.107 & 10.951 & 0.003 & 0.001 & 0.001 & 0.001 & 0.001 & 0 & 0 & 0 & 0 & 0 & -1 \\
118 & 1204 & +08:13:33.4 & -05:53:35 & 12.556 & 11.403 & 11.428 & 11.534 & 11.619 & 0.001 & 0.001 & 0.001 & 0.001 & 0.001 & 0 & 0 & 0 & 0 & 0 & -1 \\
124 & 1393 & +08:13:55.7 & -05:51:27 & 12.685 & 11.510 & 11.557 & 11.679 & 11.782 & 0.001 & 0.001 & 0.001 & 0.001 & 0.001 & 0 & 0 & 0 & 0 & 0 & 96 \\
126 & 1364 & +08:13:52.5 & -05:44:24 & 12.756 & 11.570 & 11.592 & 11.689 & 11.772 & 0.001 & 0.001 & 0.001 & 0.001 & 0.001 & 0 & 0 & 0 & 0 & 0 & 97 \\
129 & -1 & +08:13:23.9 & -05:45:13 & 12.677 & 11.766 & 11.611 & 11.482 & 11.593 & 0.001 & 0.001 & 0.001 & 0.001 & 0.001 & 0 & 0 & 0 & 0 & 0 & -1 \\
133 & 901 & +08:12:55.4 & -05:45:02 & 12.840 & 11.847 & 11.637 & 11.606 & 11.634 & 0.001 & 0.001 & 0.001 & 0.001 & 0.001 & 0 & 0 & 0 & 0 & 0 & -1 \\
136 & 1392 & +08:13:55.7 & -05:48:17 & 12.788 & 11.621 & 11.656 & 11.752 & 11.837 & 0.001 & 0.001 & 0.001 & 0.001 & 0.001 & 0 & 0 & 0 & 0 & 0 & -1 \\
141 & 1324 & +08:13:47.6 & -05:46:00 & 12.822 & 11.654 & 11.675 & 11.777 & 11.872 & 0.001 & 0.001 & 0.001 & 0.001 & 0.001 & 0 & 0 & 0 & 0 & 0 & 96 \\
143 & 1261 & +08:13:40.2 & -05:54:18 & 12.818 & 11.687 & 11.692 & 11.770 & 11.852 & 0.001 & 0.001 & 0.001 & 0.001 & 0.001 & 0 & 0 & 0 & 0 & 0 & 97 \\
151 & 1438 & +08:14:00.3 & -05:49:57 & 13.098 & 11.956 & 11.801 & 11.809 & 11.857 & 0.001 & 0.001 & 0.001 & 0.001 & 0.001 & 0 & 0 & 0 & 0 & 0 & 88 \\
158 & 2531 & +08:13:53.1 & -05:49:21 & 13.345 & 12.174 & 11.878 & 11.820 & 11.840 & 0.001 & 0.001 & 0.001 & 0.001 & 0.001 & 0 & 0 & 0 & 0 & 0 & -1 \\
161 & -1 & +08:13:33.8 & -05:53:46 & 17.615 & 11.734 & 11.917 & 15.487 & 15.284 & 0.023 & 0.001 & 0.001 & 0.011 & 0.022 & 0 & 0 & 0 & 0 & 0 & -1 \\
168 & 1129 & +08:13:24.3 & -05:44:29 & 13.243 & 12.112 & 12.010 & 12.050 & 12.100 & 0.001 & 0.001 & 0.001 & 0.001 & 0.002 & 0 & 0 & 0 & 0 & 0 & -1 \\
172 & 1265 & +08:13:40.5 & -05:52:24 & 13.331 & 12.231 & 12.089 & 12.090 & 12.142 & 0.001 & 0.001 & 0.001 & 0.001 & 0.002 & 0 & 0 & 0 & 0 & 0 & 78 \\
175 & 1474 & +08:14:06.5 & -05:46:02 & 13.320 & 12.209 & 12.113 & 12.158 & 12.228 & 0.001 & 0.001 & 0.001 & 0.001 & 0.002 & 0 & 0 & 0 & 0 & 0 & -1 \\
177 & -1 & +08:13:33.6 & -05:49:39 & 13.440 & 12.370 & 12.136 & 12.106 & 12.146 & 0.002 & 0.001 & 0.001 & 0.001 & 0.002 & 0 & 0 & 0 & 0 & 0 & -1 \\
182 & -1 & +08:13:31.8 & -05:49:29 & 13.504 & 12.459 & 12.206 & 12.163 & 12.196 & 0.002 & 0.001 & 0.001 & 0.001 & 0.002 & 0 & 0 & 0 & 0 & 0 & -1 \\
187 & 1444 & +08:14:01.9 & -05:46:51 & 13.616 & 12.520 & 12.231 & 12.166 & 12.179 & 0.002 & 0.001 & 0.001 & 0.001 & 0.002 & 0 & 0 & 0 & 0 & 0 & 94 \\
194 & -1 & +08:12:58.3 & -05:45:38 & 15.218 & 13.036 & 12.290 & 12.022 & 11.882 & 0.005 & 0.001 & 0.001 & 0.001 & 0.002 & 0 & 0 & 0 & 0 & 0 & -1 \\
\hline
\multicolumn{20}{l}{Column (1): ID numbers are ordered in increasing r' magnitude. For multiple observations of a star the r' magnitude with smallest error is used.}\\
\multicolumn{20}{l}{Colunm (2): A value of -1 indicated that there was no {\tt webda} entry matching the star's coordinates.}\\
\multicolumn{20}{l}{Column (3): DEC is listed in 2000 coordinates in HH:MM:SS.S format.}\\
\multicolumn{20}{l}{Column (4): DEC is listed in 2000 coordinates in DD:MM:SS format.}\\
\multicolumn{20}{l}{Columns (5-14): A value of -100.000 indicated either saturation or no detection}\\
\multicolumn{20}{l}{Columns (15-19):A saturation flag of 1 indicates either no detection or saturations.}\\
\multicolumn{20}{l}{Colunm (20):A membership probability of -1 indicated that there is no membership information for that star}\\
\enddata

\end{deluxetable}

\clearpage

\begin{deluxetable}{cccccccccccccccccccc}
\rotate
\tabletypesize{\tiny}
%\tabletypesize{\scriptsize}
\tablecaption{Example of Available Data for the Photometric Telescope\label{dataPT}}
\tablewidth{0pt}
\tablenum{5}
\tablehead{
  \colhead{Star ID} &
  \colhead{WEBDA ID} &
  \colhead{RA} &
  \colhead{DEC} &
  \multicolumn{5}{c}{Magnitude} &	
  \multicolumn{5}{c}{Magnitude Error} &
  \multicolumn{5}{c}{Saturation Flag} &
  \colhead{Membership } \\	
  \colhead{} &
 \colhead{} &	 
  \colhead{} &
  \colhead{} &
  \colhead{ $u'$ } &
  \colhead{ $g'$ } &
  \colhead{ $r'$ } &
  \colhead{ $i'$ } &
  \colhead{ $z'$ } &
 \colhead{ $u'$ } &
  \colhead{ $g'$ } &
  \colhead{ $r'$ } &
  \colhead{ $i'$ } &
  \colhead{ $z'$ } &
 \colhead{ $u'$ } &
  \colhead{ $g'$ } &
  \colhead{ $r'$ } &
  \colhead{ $i'$ } &
  \colhead{ $z'$ } &
  \colhead{Probability}\\
 \colhead{1} &
 \colhead{2} &
 \colhead{3} &
 \colhead{4} &
 \colhead{5} &
 \colhead{6} &
 \colhead{7} &
 \colhead{8} & 
 \colhead{9} &
 \colhead{10} &
 \colhead{11} &
 \colhead{12} &
 \colhead{13} &
 \colhead{14} &
 \colhead{15} &
 \colhead{16} &
 \colhead{17} &
 \colhead{18} &
 \colhead{19} &
 \colhead{20} 
}
\startdata
9  &  1241  &  +08:13:38.0  &  -06:01:32  &  12.741  &  10.062  &  9.150  &  8.819  &  8.662  &  0.006  &  0.003  &  0.004  &  0.005  &  0.001  &  0  &  0  &  0  &  0  &  0  &  -1 \\
34  &  1367  &  +08:13:52.9  &  -05:42:46  &  10.342  &  -100.000  &  9.490  &  9.672  &  9.790  &  0.002  &  -100.000  &  0.004  &  0.005  &  0.002  &  0  &  1  &  0  &  0  &  0  &  97 \\
22  &  1616  &  +08:14:26.3  &  -05:44:34  &  12.886  &  10.341  &  9.493  &  9.188  &  9.042  &  0.006  &  0.003  &  0.004  &  0.005  &  0.001  &  0  &  0  &  0  &  0  &  0  &  2 \\
29  &  1521  &  +08:14:12.5  &  -05:33:57  &  10.743  &  9.702  &  9.504  &  9.461  &  9.501  &  0.002  &  0.003  &  0.004  &  0.005  &  0.002  &  0  &  0  &  0  &  0  &  0  &  0 \\
25  &  978  &  +08:13:05.3  &  -05:45:00  &  10.391  &  9.324  &  9.516  &  9.711  &  9.841  &  0.002  &  0.003  &  0.004  &  0.005  &  0.002  &  0  &  0  &  0  &  0  &  0  &  96 \\
28  &  1169  &  +08:13:28.6  &  -05:48:15  &  10.427  &  9.382  &  9.568  &  9.761  &  9.896  &  0.002  &  0.003  &  0.004  &  0.005  &  0.002  &  0  &  0  &  0  &  0  &  0  &  96 \\
32  &  920  &  +08:12:58.5  &  -05:34:08  &  10.508  &  9.394  &  9.569  &  9.752  &  9.870  &  0.002  &  0.003  &  0.004  &  0.005  &  0.002  &  0  &  0  &  0  &  0  &  0  &  92 \\
40  &  1765  &  +08:14:48.5  &  -05:48:43  &  10.996  &  9.801  &  9.787  &  9.885  &  9.988  &  0.002  &  0.003  &  0.004  &  0.005  &  0.002  &  0  &  0  &  0  &  0  &  0  &  0 \\
37  &  1183  &  +08:13:30.4  &  -06:04:04  &  10.901  &  9.751  &  9.798  &  9.915  &  10.022  &  0.002  &  0.003  &  0.004  &  0.005  &  0.002  &  0  &  0  &  0  &  0  &  0  &  -1 \\
38  &  1253  &  +08:13:39.6  &  -05:47:14  &  10.806  &  9.689  &  9.798  &  9.946  &  10.035  &  0.002  &  0.003  &  0.004  &  0.005  &  0.002  &  0  &  0  &  0  &  0  &  0  &  95 \\
39  &  1117  &  +08:13:23.1  &  -05:45:23  &  10.640  &  9.622  &  9.826  &  10.030  &  10.157  &  0.002  &  0.003  &  0.004  &  0.005  &  0.002  &  0  &  0  &  0  &  0  &  0  &  97 \\
41  &  1434  &  +08:13:59.6  &  -05:53:23  &  14.364  &  11.046  &  9.886  &  9.423  &  9.178  &  0.017  &  0.003  &  0.004  &  0.005  &  0.002  &  0  &  0  &  0  &  0  &  0  &  41 \\
42  &  1600  &  +08:14:24.1  &  -05:31:15  &  12.483  &  10.567  &  9.912  &  9.678  &  9.604  &  0.005  &  0.003  &  0.004  &  0.005  &  0.002  &  0  &  0  &  0  &  0  &  0  &  -1 \\
45  &  1338  &  +08:13:48.9  &  -05:44:23  &  10.750  &  9.760  &  9.960  &  10.148  &  10.279  &  0.002  &  0.003  &  0.004  &  0.005  &  0.002  &  0  &  0  &  0  &  0  &  0  &  96 \\
44  &  975  &  +08:13:04.9  &  -05:53:04  &  10.877  &  9.799  &  9.964  &  10.149  &  10.274  &  0.002  &  0.003  &  0.004  &  0.005  &  0.002  &  0  &  0  &  0  &  0  &  0  &  89 \\
46  &  1642  &  +08:14:30.7  &  -05:46:53  &  11.207  &  10.113  &  9.996  &  10.005  &  10.070  &  0.002  &  0.003  &  0.004  &  0.005  &  0.002  &  0  &  0  &  0  &  0  &  0  &  0 \\
50  &  1147  &  +08:13:26.5  &  -05:49:54  &  10.863  &  9.833  &  10.041  &  10.238  &  10.384  &  0.002  &  0.003  &  0.004  &  0.005  &  0.002  &  0  &  0  &  0  &  0  &  0  &  95 \\
51  &  1281  &  +08:13:43.3  &  -05:41:33  &  10.840  &  9.826  &  10.043  &  10.251  &  10.378  &  0.002  &  0.003  &  0.004  &  0.005  &  0.002  &  0  &  0  &  0  &  0  &  0  &  97 \\
56  &  1256  &  +08:13:40.3  &  -05:42:20  &  11.105  &  10.002  &  10.159  &  10.327  &  10.451  &  0.002  &  0.003  &  0.004  &  0.005  &  0.002  &  0  &  0  &  0  &  0  &  0  &  97 \\
58  &  1382  &  +08:13:54.3  &  -05:58:47  &  11.049  &  10.023  &  10.221  &  10.418  &  10.552  &  0.002  &  0.003  &  0.004  &  0.005  &  0.002  &  0  &  0  &  0  &  0  &  0  &  96 \\
60  &  1453  &  +08:14:03.1  &  -05:41:44  &  11.011  &  10.005  &  10.227  &  10.433  &  10.575  &  0.002  &  0.003  &  0.004  &  0.005  &  0.002  &  0  &  0  &  0  &  0  &  0  &  94 \\
57  &  1529  &  +08:14:13.7  &  -05:27:20  &  12.326  &  10.826  &  10.248  &  10.066  &  9.976  &  0.004  &  0.003  &  0.004  &  0.005  &  0.002  &  0  &  0  &  0  &  0  &  0  &  -1 \\
61  &  1306  &  +08:13:45.9  &  -05:46:02  &  11.266  &  10.216  &  10.299  &  10.432  &  10.534  &  0.002  &  0.003  &  0.004  &  0.005  &  0.002  &  0  &  0  &  0  &  0  &  0  &  -1 \\
62  &  1029  &  +08:13:12.1  &  -05:46:42  &  11.315  &  10.229  &  10.360  &  10.515  &  10.629  &  0.003  &  0.003  &  0.004  &  0.005  &  0.002  &  0  &  0  &  0  &  0  &  0  &  96 \\
63  &  1337  &  +08:13:48.2  &  -06:05:43  &  13.751  &  11.283  &  10.452  &  10.159  &  10.000  &  0.011  &  0.003  &  0.004  &  0.005  &  0.002  &  0  &  0  &  0  &  0  &  0  &  10 \\
64  &  -1  &  +08:13:19.6  &  -05:33:37  &  11.362  &  10.271  &  10.467  &  -100.000  &  10.600  &  0.003  &  0.003  &  0.004  &  -100.000  &  0.002  &  0  &  0  &  0  &  1  &  0  &  -1 \\
52  &  -1  &  +08:13:43.1  &  -05:45:53  &  11.343  &  10.285  &  10.497  &  10.679  &  10.759  &  0.003  &  0.003  &  0.004  &  0.005  &  0.002  &  0  &  0  &  0  &  0  &  0  &  -1 \\
65  &  848  &  +08:12:48.5  &  -05:53:44  &  11.743  &  10.625  &  10.521  &  10.564  &  10.637  &  0.003  &  0.003  &  0.004  &  0.005  &  0.002  &  0  &  0  &  0  &  0  &  0  &  52 \\
66  &  -1  &  +08:13:26.4  &  -05:42:34  &  11.867  &  10.805  &  10.537  &  10.484  &  10.499  &  0.003  &  0.003  &  0.004  &  0.005  &  0.002  &  0  &  0  &  0  &  0  &  0  &  -1 \\
68  &  1069  &  +08:13:17.5  &  -05:41:13  &  11.449  &  10.377  &  10.558  &  10.747  &  10.877  &  0.003  &  0.003  &  0.004  &  0.005  &  0.003  &  0  &  0  &  0  &  0  &  0  &  87 \\
69  &  1576  &  +08:14:20.2  &  -05:39:57  &  11.435  &  10.375  &  10.581  &  10.772  &  10.914  &  0.003  &  0.003  &  0.004  &  0.005  &  0.003  &  0  &  0  &  0  &  0  &  0  &  -1 \\
72  &  1362  &  +08:13:52.0  &  -05:54:20  &  11.605  &  10.496  &  10.595  &  10.734  &  10.843  &  0.003  &  0.003  &  0.004  &  0.005  &  0.003  &  0  &  0  &  0  &  0  &  0  &  91 \\
70  &  1388  &  +08:13:55.0  &  -05:57:58  &  12.388  &  11.053  &  10.604  &  10.473  &  10.461  &  0.005  &  0.003  &  0.004  &  0.005  &  0.002  &  0  &  0  &  0  &  0  &  0  &  -1 \\
71  &  1449  &  +08:14:02.3  &  -05:56:47  &  11.545  &  10.440  &  10.604  &  10.797  &  10.966  &  0.003  &  0.003  &  0.004  &  0.005  &  0.003  &  0  &  0  &  0  &  0  &  0  &  96 \\
76  &  1276  &  +08:13:42.8  &  -05:35:47  &  11.508  &  10.477  &  10.636  &  10.837  &  10.973  &  0.003  &  0.003  &  0.004  &  0.005  &  0.003  &  0  &  0  &  0  &  0  &  0  &  95 \\
75  &  1042  &  +08:13:13.7  &  -05:56:38  &  11.564  &  10.488  &  10.657  &  10.836  &  10.978  &  0.003  &  0.003  &  0.004  &  0.005  &  0.003  &  0  &  0  &  0  &  0  &  0  &  88 \\
77  &  1406  &  +08:13:57.3  &  -05:48:15  &  11.601  &  10.504  &  10.667  &  10.847  &  10.977  &  0.003  &  0.003  &  0.004  &  0.005  &  0.003  &  0  &  0  &  0  &  0  &  0  &  94 \\
79  &  1424  &  +08:13:58.9  &  -05:50:40  &  11.834  &  10.676  &  10.719  &  10.844  &  10.952  &  0.003  &  0.003  &  0.004  &  0.005  &  0.003  &  0  &  0  &  0  &  0  &  0  &  96 \\
83  &  1677  &  +08:14:36.5  &  -05:45:53  &  11.865  &  10.758  &  10.804  &  10.910  &  11.012  &  0.003  &  0.003  &  0.004  &  0.005  &  0.003  &  0  &  0  &  0  &  0  &  0  &  -1 \\
82  &  1632  &  +08:14:28.0  &  -06:02:46  &  11.869  &  10.731  &  10.808  &  10.941  &  11.050  &  0.003  &  0.003  &  0.004  &  0.005  &  0.003  &  0  &  0  &  0  &  0  &  0  &  18 \\
81  &  1454  &  +08:14:02.9  &  -05:50:27  &  12.404  &  11.199  &  10.811  &  10.692  &  10.681  &  0.005  &  0.003  &  0.004  &  0.005  &  0.002  &  0  &  0  &  0  &  0  &  0  &  0 \\
84  &  1214  &  +08:13:34.6  &  -05:55:37  &  13.963  &  11.631  &  10.830  &  10.556  &  10.414  &  0.013  &  0.004  &  0.004  &  0.005  &  0.002  &  0  &  0  &  0  &  0  &  0  &  0 \\
87  &  1694  &  +08:14:39.1  &  -05:51:47  &  11.793  &  10.694  &  10.850  &  11.023  &  11.144  &  0.003  &  0.003  &  0.004  &  0.005  &  0.003  &  0  &  0  &  0  &  0  &  0  &  88 \\
88  &  1470  &  +08:14:05.3  &  -05:56:32  &  11.799  &  10.714  &  10.876  &  11.050  &  11.189  &  0.003  &  0.003  &  0.004  &  0.005  &  0.003  &  0  &  0  &  0  &  0  &  0  &  -1 \\
89  &  1586  &  +08:14:21.7  &  -05:47:23  &  12.068  &  10.944  &  10.938  &  11.002  &  11.095  &  0.004  &  0.003  &  0.004  &  0.005  &  0.003  &  0  &  0  &  0  &  0  &  0  &  97 \\
86  &  1230  &  +08:13:36.6  &  -05:51:23  &  14.833  &  11.930  &  10.943  &  10.577  &  10.498  &  0.025  &  0.004  &  0.004  &  0.005  &  0.002  &  0  &  0  &  0  &  0  &  0  &  57 \\
96  &  1458  &  +08:14:04.1  &  -05:28:45  &  12.184  &  11.101  &  10.968  &  10.964  &  11.017  &  0.004  &  0.003  &  0.004  &  0.005  &  0.003  &  0  &  0  &  0  &  0  &  0  &  -1 \\
92  &  1804  &  +08:14:55.2  &  -05:34:45  &  13.590  &  11.632  &  10.969  &  10.739  &  10.637  &  0.010  &  0.004  &  0.004  &  0.005  &  0.002  &  0  &  0  &  0  &  0  &  0  &  -1 \\
93  &  -1  &  +08:13:40.9  &  -05:46:57  &  12.025  &  10.897  &  10.976  &  11.108  &  11.208  &  0.004  &  0.003  &  0.004  &  0.005  &  0.003  &  0  &  0  &  0  &  0  &  0  &  87 \\
95  &  924  &  +08:12:58.4  &  -05:51:06  &  12.065  &  10.946  &  10.984  &  11.100  &  11.217  &  0.004  &  0.003  &  0.004  &  0.005  &  0.003  &  0  &  0  &  0  &  0  &  0  &  95 \\
\hline
\multicolumn{20}{l}{Column (1): ID numbers are ordered in increasing r' magnitude. For multiple observations of a star the r' magnitude with smallest error is used.}\\
\multicolumn{20}{l}{Colunm (2): A value of -1 indicated that there was no {\tt webda} entry matching the star's coordinates.}\\
\multicolumn{20}{l}{Column (3): DEC is listed in 2000 coordinates in HH:MM:SS.S format.}\\
\multicolumn{20}{l}{Column (4): DEC is listed in 2000 coordinates in DD:MM:SS format.}\\
\multicolumn{20}{l}{Columns (5-14): A value of -100.000 indicated either saturation or no detection}\\
\multicolumn{20}{l}{Columns (15-19):A saturation flag of 1 indicates either no detection or saturations.}\\
\multicolumn{20}{l}{Colunm (20):A membership probability of -1 indicated that there is no membership information for that star}\\

\enddata

\end{deluxetable}

\clearpage

\begin{deluxetable}{lccccccccccccccccccl}
\rotate
\tablenum{6}
\tabletypesize{\tiny}
%\tabletypesize{\scriptsize}
\tablecaption{Example of Available Data for the CTIO Curtis-Schmidt Telescope\label{dataCTIO}}
\tablewidth{0pt}
\tablehead{
  \colhead{Star ID} &
  \colhead{WEBDA ID} &
  \colhead{RA} &
  \colhead{DEC} &
  \multicolumn{5}{c}{Magnitude} &	
  \multicolumn{5}{c}{Magnitude Error} &
  \multicolumn{5}{c}{Saturation Flag} &
  \colhead{Membership } \\	
  \colhead{} &
 \colhead{} &	 
  \colhead{} &
  \colhead{} &
  \colhead{ $u'$ } &
  \colhead{ $g'$ } &
  \colhead{ $r'$ } &
  \colhead{ $i'$ } &
  \colhead{ $z'$ } &
 \colhead{ $u'$ } &
  \colhead{ $g'$ } &
  \colhead{ $r'$ } &
  \colhead{ $i'$ } &
  \colhead{ $z'$ } &
 \colhead{ $u'$ } &
  \colhead{ $g'$ } &
  \colhead{ $r'$ } &
  \colhead{ $i'$ } &
  \colhead{ $z'$ } &
  \colhead{Probability}\\
 \colhead{1} &
 \colhead{2} &
 \colhead{3} &
 \colhead{4} &
 \colhead{5} &
 \colhead{6} &
 \colhead{7} &
 \colhead{8} & 
 \colhead{9} &
 \colhead{10} &
 \colhead{11} &
 \colhead{12} &
 \colhead{13} &
 \colhead{14} &
 \colhead{15} &
 \colhead{16} &
 \colhead{17} &
 \colhead{18} &
 \colhead{19} &
 \colhead{20} 
}
\startdata

1 & 1560 & +08:14:17.0 & -05:54:00 & 11.071 & 8.613 & 7.801 & 7.435 & 7.234 & 0.005 & 0.001 & 0.001 & 0.001 & 0.001 & 0 & 0 & 0 & 0 & 0 & 97 \\
2 & 1073 & +08:13:17.9 & -05:38:26 & 9.629 & 8.536 & 8.714 & 8.884 & 8.968 & 0.002 & 0.001 & 0.001 & 0.002 & 0.003 & 0 & 0 & 0 & 0 & 0 & 96 \\
3 & 1260 & +08:13:40.4 & -05:46:25 & 10.548 & 9.146 & 8.811 & 8.707 & 8.631 & 0.004 & 0.002 & 0.001 & 0.001 & 0.002 & 0 & 0 & 0 & 0 & 0 & 97 \\
4 & 1296 & +08:13:44.8 & -05:48:00 & 11.166 & 9.446 & 8.915 & 8.740 & 8.653 & 0.005 & 0.002 & 0.001 & 0.002 & 0.002 & 0 & 0 & 0 & 0 & 0 & 96 \\
5 & 776 & +08:12:37.1 & -05:40:50 & 11.690 & 9.712 & 9.022 & 8.822 & 8.711 & 0.006 & 0.002 & 0.001 & 0.002 & 0.002 & 0 & 0 & 0 & 0 & 0 & 95 \\
6 & 1628 & +08:14:28.1 & -05:42:16 & 12.052 & 9.823 & 9.063 & 8.823 & 8.679 & 0.007 & 0.002 & 0.001 & 0.002 & 0.002 & 0 & 0 & 0 & 0 & 0 & 96 \\
8 & -1 & +08:14:17.2 & -05:54:00 & 11.056 & 9.182 & 9.101 & 8.604 & 8.423 & 0.002 & 0.001 & 0.001 & 0.001 & 0.001 & 0 & 0 & 0 & 0 & 0 & -1 \\
7 & 1313 & +08:13:46.6 & -05:44:52 & 10.010 & 8.918 & 9.102 & 9.267 & 9.375 & 0.003 & 0.001 & 0.001 & 0.002 & 0.003 & 0 & 0 & 0 & 0 & 0 & 5 \\
9 & 1241 & +08:13:38.0 & -06:01:32 & 12.707 & 10.066 & 9.136 & 8.810 & 8.614 & 0.010 & 0.002 & 0.002 & 0.002 & 0.002 & 0 & 0 & 0 & 0 & 0 & 59 \\
10 & 1832 & +08:15:00.1 & -05:32:04 & 10.894 & 9.589 & 9.182 & 9.086 & 8.995 & 0.002 & 0.002 & 0.002 & 0.001 & 0.001 & 0 & 0 & 0 & 0 & 0 & 0 \\
12 & 870 & +08:12:51.2 & -05:50:50 & 12.449 & 10.101 & 9.242 & 8.936 & 8.750 & 0.009 & 0.002 & 0.002 & 0.002 & 0.002 & 0 & 0 & 0 & 0 & 0 & 87 \\
14 & 1005 & +08:13:08.5 & -05:38:35 & 10.092 & 9.046 & 9.278 & 9.457 & 9.561 & 0.003 & 0.001 & 0.002 & 0.002 & 0.003 & 0 & 0 & 0 & 0 & 0 & 97 \\
15 & 1320 & +08:13:47.6 & -05:37:25 & 9.845 & 9.275 & 9.424 & 9.229 & 9.206 & 0.001 & 0.001 & 0.001 & 0.001 & 0.001 & 0 & 0 & 0 & 0 & 0 & 88 \\
16 & 1260 & +08:13:40.5 & -05:46:25 & 10.553 & 9.414 & 9.428 & 9.049 & 8.670 & 0.001 & 0.001 & 0.001 & 0.001 & 0.001 & 0 & 0 & 0 & 0 & 0 & -1 \\
17 & -1 & +08:13:45.0 & -05:48:00 & 11.168 & 9.600 & 9.468 & 9.316 & 8.655 & 0.002 & 0.001 & 0.001 & 0.001 & 0.001 & 0 & 0 & 0 & 0 & 0 & -1 \\
18 & 773 & +08:12:36.4 & -05:39:50 & 13.135 & 10.253 & 9.482 & 9.149 & 8.623 & 0.004 & 0.001 & 0.001 & 0.001 & 0.001 & 0 & 0 & 0 & 0 & 0 & 0 \\
19 & 1541 & +08:14:15.4 & -05:43:15 & 10.120 & 9.459 & 9.490 & 9.323 & 9.366 & 0.001 & 0.001 & 0.001 & 0.001 & 0.001 & 0 & 0 & 0 & 0 & 0 & 95 \\
20 & 1241 & +08:13:38.1 & -06:01:31 & 12.705 & 10.059 & 9.493 & 9.114 & 8.599 & 0.004 & 0.001 & 0.001 & 0.001 & 0.001 & 0 & 0 & 0 & 0 & 0 & 59 \\
21 & -1 & +08:14:16.2 & -05:54:00 & 11.042 & 10.018 & 9.496 & 8.604 & 8.400 & 0.002 & 0.001 & 0.001 & 0.001 & 0.001 & 0 & 0 & 0 & 0 & 0 & -1 \\
22 & 1616 & +08:14:26.3 & -05:44:34 & 12.906 & 10.358 & 9.499 & 9.206 & 9.014 & 0.011 & 0.003 & 0.002 & 0.002 & 0.003 & 0 & 0 & 0 & 0 & 0 & 2 \\
13 & 1218 & +08:13:35.4 & -05:53:02 & 11.797 & 9.912 & 9.503 & 9.070 & 8.925 & 0.002 & 0.001 & 0.001 & 0.001 & 0.001 & 0 & 0 & 0 & 0 & 0 & 95 \\
23 & -1 & +08:14:28.3 & -05:42:16 & 12.066 & 9.825 & 9.511 & 9.046 & 8.682 & 0.003 & 0.001 & 0.001 & 0.001 & 0.001 & 0 & 0 & 0 & 0 & 0 & -1 \\
11 & 1289 & +08:13:44.2 & -05:48:48 & 9.991 & 9.383 & 9.513 & 9.413 & 9.521 & 0.001 & 0.001 & 0.001 & 0.001 & 0.001 & 0 & 0 & 0 & 0 & 0 & 93 \\
24 & 1313 & +08:13:46.8 & -05:44:52 & 9.980 & 9.940 & 9.518 & 9.289 & 9.356 & 0.001 & 0.001 & 0.001 & 0.001 & 0.001 & 0 & 0 & 0 & 0 & 0 & -1 \\
26 & -1 & +08:13:17.6 & -05:38:26 & 9.658 & 9.281 & 9.540 & 9.031 & 8.981 & 0.001 & 0.001 & 0.001 & 0.001 & 0.001 & 0 & 0 & 0 & 0 & 0 & -1 \\
27 & -1 & +08:12:37.4 & -05:40:50 & 11.690 & 9.817 & 9.559 & 9.309 & 8.730 & 0.002 & 0.001 & 0.001 & 0.001 & 0.001 & 0 & 0 & 0 & 0 & 0 & -1 \\
29 & 1521 & +08:14:12.5 & -05:33:57 & 10.768 & 9.729 & 9.639 & 9.491 & 9.498 & 0.001 & 0.001 & 0.001 & 0.001 & 0.003 & 0 & 0 & 0 & 0 & 0 & 0 \\
30 & 870 & +08:12:51.4 & -05:50:50 & 12.469 & 10.118 & 9.642 & 9.055 & 8.775 & 0.003 & 0.001 & 0.001 & 0.001 & 0.001 & 0 & 0 & 0 & 0 & 0 & 87 \\
31 & 1616 & +08:14:26.5 & -05:44:34 & 12.886 & 10.359 & 9.675 & 9.264 & 9.013 & 0.004 & 0.001 & 0.001 & 0.001 & 0.001 & 0 & 0 & 0 & 0 & 0 & 2 \\
32 & 920 & +08:12:58.5 & -05:34:08 & 10.526 & 9.562 & 9.676 & 9.797 & 9.888 & 0.001 & 0.001 & 0.001 & 0.001 & 0.002 & 0 & 0 & 0 & 0 & 0 & 92 \\
28 & 1169 & +08:13:28.6 & -05:48:15 & 10.425 & 9.604 & 9.678 & 9.774 & 9.888 & 0.001 & 0.001 & 0.001 & 0.001 & 0.002 & 0 & 0 & 0 & 0 & 0 & 96 \\
33 & -1 & +08:13:08.7 & -05:38:35 & 10.102 & 9.703 & 9.679 & 9.482 & 9.566 & 0.001 & 0.001 & 0.001 & 0.001 & 0.001 & 0 & 0 & 0 & 0 & 0 & 97 \\
34 & 1367 & +08:13:52.9 & -05:42:46 & 10.341 & 9.482 & 9.727 & 9.696 & 9.804 & 0.001 & 0.001 & 0.001 & 0.001 & 0.002 & 0 & 0 & 0 & 0 & 0 & 97 \\
25 & 978 & +08:13:05.3 & -05:45:00 & 10.401 & 9.577 & 9.746 & 9.736 & 9.823 & 0.001 & 0.001 & 0.001 & 0.001 & 0.002 & 0 & 0 & 0 & 0 & 0 & 96 \\
35 & -1 & +08:12:36.7 & -05:40:50 & 11.690 & 9.820 & 9.756 & 9.309 & 8.730 & 0.002 & 0.001 & 0.001 & 0.001 & 0.001 & 0 & 0 & 0 & 0 & 0 & -1 \\
36 & -1 & +08:13:18.6 & -05:38:26 & 9.645 & 10.048 & 9.763 & -100.000 & 8.973 & 0.001 & 0.001 & 0.001 & -100.000 & 0.001 & 0 & 0 & 0 & 1 & 0 & -1 \\
38 & 1183 & +08:13:30.4 & -06:04:04 & 10.889 & 9.769 & 9.810 & 9.900 & 9.963 & 0.002 & 0.001 & 0.001 & 0.001 & 0.002 & 0 & 0 & 0 & 0 & 0 & -1 \\
40 & 1765 & +08:14:48.5 & -05:48:43 & 11.009 & 9.808 & 9.839 & 9.890 & 9.965 & 0.002 & 0.001 & 0.001 & 0.001 & 0.002 & 0 & 0 & 0 & 0 & 0 & 0 \\
37 & 1253 & +08:13:39.6 & -05:47:14 & 10.796 & 9.699 & 9.881 & 9.970 & 10.042 & 0.001 & 0.001 & 0.001 & 0.001 & 0.002 & 0 & 0 & 0 & 0 & 0 & 95 \\
41 & 1434 & +08:13:59.6 & -05:53:22 & 14.351 & 11.063 & 9.883 & 9.420 & 9.148 & 0.008 & 0.002 & 0.001 & 0.001 & 0.001 & 0 & 0 & 0 & 0 & 0 & 41 \\
39 & 1117 & +08:13:23.0 & -05:45:23 & -100.000 & 9.708 & 9.885 & -100.000 & -100.000 & -100.000 & 0.001 & 0.001 & -100.000 & -100.000 & 1 & 0 & 0 & 1 & 1 & -1 \\
42 & 1600 & +08:14:24.1 & -05:31:15 & 12.515 & 10.583 & 9.913 & 9.697 & 9.590 & 0.003 & 0.001 & 0.001 & 0.001 & 0.001 & 0 & 0 & 0 & 0 & 0 & -1 \\
43 & 1726 & +08:14:43.2 & -05:27:07 & 10.783 & 9.725 & 9.941 & 10.110 & 10.214 & 0.001 & 0.001 & 0.001 & 0.001 & 0.002 & 0 & 0 & 0 & 0 & 0 & -1 \\
45 & 1338 & +08:13:48.9 & -05:44:23 & 10.775 & 9.795 & 9.992 & 10.172 & 10.278 & 0.001 & 0.001 & 0.001 & 0.001 & 0.002 & 0 & 0 & 0 & 0 & 0 & 96 \\
46 & 1642 & +08:14:30.7 & -05:46:53 & 11.209 & 10.111 & 9.995 & 10.006 & 10.028 & 0.002 & 0.001 & 0.001 & 0.001 & 0.002 & 0 & 0 & 0 & 0 & 0 & 0 \\
44 & 975 & +08:13:04.9 & -05:53:04 & 10.883 & 9.818 & 10.011 & 10.177 & 10.252 & 0.002 & 0.001 & 0.001 & 0.001 & 0.002 & 0 & 0 & 0 & 0 & 0 & 89 \\
47 & -1 & +08:14:14.7 & -05:54:01 & -100.000 & -100.000 & 10.027 & 9.878 & -100.000 & -100.000 & -100.000 & 0.001 & 0.001 & -100.000 & 1 & 1 & 0 & 0 & 1 & -1 \\
48 & 1527 & +08:14:12.0 & -06:09:08 & 10.981 & 9.879 & 10.037 & 10.174 & 10.291 & 0.002 & 0.001 & 0.001 & 0.001 & 0.002 & 0 & 0 & 0 & 0 & 0 & -1 \\
50 & 1147 & +08:13:26.5 & -05:49:53 & 10.869 & 9.846 & 10.067 & 10.261 & 10.369 & 0.002 & 0.001 & 0.001 & 0.001 & 0.002 & 0 & 0 & 0 & 0 & 0 & 95 \\
52 & 1281 & +08:13:43.3 & -05:41:33 & 10.857 & 9.837 & 10.077 & 10.265 & 10.383 & 0.002 & 0.001 & 0.001 & 0.001 & 0.002 & 0 & 0 & 0 & 0 & 0 & 97 \\
\hline
\multicolumn{20}{l}{Column (1): ID numbers are ordered in increasing r' magnitude. For multiple observations of a star the r' magnitude with smallest error is used.}\\
\multicolumn{20}{l}{Colunm (2): A value of -1 indicated that there was no {\tt webda} entry matching the star's coordinates.}\\
\multicolumn{20}{l}{Column (3): DEC is listed in 2000 coordinates in HH:MM:SS.S format.}\\
\multicolumn{20}{l}{Column (4): DEC is listed in 2000 coordinates in DD:MM:SS format.}\\
\multicolumn{20}{l}{Columns (5-14): A value of -100.000 indicated either saturation or no detection}\\
\multicolumn{20}{l}{Columns (15-19):A saturation flag of 1 indicates either no detection or saturations.}\\
\multicolumn{20}{l}{Colunm (20):A membership probability of -1 indicated that there is no membership information for that star}\\
\enddata

\end{deluxetable}

\end{document}